\documentclass[%
 reprint,
superscriptaddress,
 amsmath,amssymb,
 aps,prc
]{revtex4-1}
\usepackage{mathtools,amssymb,lipsum}
\usepackage{dcolumn}
\usepackage{bm}
\usepackage{amsmath}

\begin{document}

\preprint{APS/123-QED}

\title{Accurately accounting for effects on times-of-flight caused by \\finite field-transition times during the ejection of ions from a storage trap:\\ A study for single-reference TOF and MRTOF mass spectrometry}

\author{M. Rosenbusch}
\altaffiliation{Present address: Wako Nuclear Science Center (WNSC), Institute of Particle and Nuclear Studies (IPNS), High Energy Accelerator Research Organization (KEK), Wako, Saitama 351-0198, Japan}
\affiliation{%
 RIKEN Nishina Center for Accelerator-Based Science, Wako, Saitama 351-0198, Japan\\
}%
\author{P. Schury}%
\affiliation{%
 Wako Nuclear Science Center (WNSC), Institute of Particle and Nuclear Studies (IPNS), High Energy Accelerator Research Organization (KEK), Wako, Saitama 351-0198, Japan\\
}%
\author{M. Wada}%
\affiliation{%
 Wako Nuclear Science Center (WNSC), Institute of Particle and Nuclear Studies (IPNS), High Energy Accelerator Research Organization (KEK), Wako, Saitama 351-0198, Japan\\
}%
\author{S. Iimura}
\affiliation{%
 Department of Physics, Graduate School of Science, Osaka University, 1-1 Machikaneyama, Toyonaka, Osaka 560-0043, Japan\\
}%
\affiliation{%
 RIKEN nishina center for accelerator-based science, Wako, Saitama 351-0198, Japan\\
}%
\author{Y. Ito}
\affiliation{%
	Japanese Atomic Energy Agency (JAEA), Ibaraki 319-1195, Japan\\
}%
\author{H. Wollnik}%
\affiliation{%
 New Mexico State University, Las Cruces, NM 88001, USA\\
}%

\date{\today}

\begin{abstract}
	In applied forms of time-of-flight mass spectrometry utilizing ion storage devices prior to an analysis device, a non instantaneous electric ejection pulse applied in the region of ion storage is used to accelerate ions into the time-of-flight analyzer. The calculated mass value of the ions from the time-of-flight is dependent on the duration of the field transition up to full strength. For novel applications dedicated to precision measurements, such as multi-reflection time-of-flight mass spectrometry of short-lived isotopes, the goal is to continuously decrease the measurement uncertainty while providing a mass accuracy on the same order. Even though dynamic-field models for time-of-flight mass spectrometry have been considered in the past for technological advances, it is important to study the accuracy of the measured mass in this context. Using a simplified linear model for the field transition, we provide a basic investigation of the scenario, and discuss the deviation from the classical "mass-over-charge" dependency of the ions' time-of-flight, which becomes violated. The emerging mass discrepancy depends on the distance between the mass of the ion used for calibration and that of the ion of interest and, in extreme cases, can increase to about one percent for systems with short times-of-flight. However, for typical conditions in single-reference multi-reflection time-of-flight mass spectrometry, mass deviations caused by this effect typically remain below the $1\,\mathrm{ppm}$ level. If a mass calibration using two or more ion species is possible during the measurement, the effect becomes negligible for appropriate choices of reference masses.
\end{abstract}

\maketitle


\section{\label{Intro}Introduction}

Time-of-flight mass spectrometry (TOF-MS) technology \cite[]{Wolff1953,Wiley1955,Katzenstein1955,Mamyrin1973} has become one of today's basic tools for particle identification and mass measurements in analytic chemistry, medicine, biology, and many fields of physics. Ions are created or stored at a chosen starting position and subsequently accelerated and guided by static electric fields (see also approaches using static magnetic fields \cite[]{Goudsmit1948,Hays1951}) until their time-of-flight is detected by impact on a detection device at the end of the intended flight path. In many TOF-MS systems used in science and industry, ions are created inside of a previously existing static electric field, and are thus instantaneously accelerated towards an ion-TOF detector \cite[]{MAMYRIN2001251}. However, in other applications, such as distance-of-flight mass spectrometry with constant-momentum acceleration \cite[]{Enke2007} or delayed ion extraction for MALDI TOF-MS \cite[]{Vestal1995} a purposely time-dependent acceleration scheme is applied to improve the performance and mass resolution. The ion motion using such acceleration schemes differs from that of the static case and the mass accuracy must be reconfirmed \cite[]{Takach1997}. Recent applications exploit the advantages of quadrupole ion traps as a preparatory stage for the ions \cite{Steven1992}, where an extraction field must be switched on to accelerate ions towards a mass spectrometer. In this case a time-dependent acceleration scheme is not intended, but unavoidable.\par
In the growing community for precision mass measurements of unstable nuclei using multi-reflection time-of-flight (MRTOF) mass spectrographs \cite[]{WOLLNIK1990267,Wienholtz2013,ITO2013,Wolf2013a,SCHURY201439,CHAUVEAU2016211,HIRSH2016229,Ito2017,Jesh2017,SanAndresUndChristine2019}, the usage of ion traps is essential for accumulation and cooling of ions prior to the injection into an MRTOF system. However, in an ion trap the stored ions are first confined inside a certain volume by a trapping potential of typically a few volts depth realized by the principles of Paul traps or Penning traps, and only after a chosen storage time the potential changes to a potential gradient of several tens to hundreds of volt per millimeter in order to extract the ions. In realistic systems, the duration of the extraction pulse can be in the order of $10\,\mathrm{ns}$ to $1\,\mathrm{\mu s}$, which depends on the electronic components used. Masses are typically calculated using the assumption of a negligible switching time, \textit{i.e} using the static-field approach, where the TOF is proportional to the charge-over-mass ratio.\par
Today TOF-MS and in particular MRTOF-MS aims to reach new levels of precision with relative mass uncertainties as low as $\delta m / m = 1\times 10^{-7}$ or even lower, as recently reported \cite[]{KIMURA2018134}. In order to investigate the robustness of the electrostatic-field approach for high levels of measurement precision, we have performed a basic investigation of the time-of-flight for a single ion in a TOF-MS system including a finite transition time of the ejection pulse from a trapping/starting region. We will point out the differences of the mass results obtained from the electrostatic theory and those obtained from our new model, which properly accounts for the dynamic acceleration field. In addition, calculations of a reflectron system \cite[]{Mamyrin1973} have been performed to benchmark the model under more realistic conditions, meaning a focusing of ions with approximate second-order corrections. The given number examples are adapted to masses of interest for the nuclear physics community, where experiments are dominantly performed using singly charged or doubly charged ions up to mass number $A\approx250$. Further, the results and the impact on the obtained mass value will be regarded for the millisecond long flight times applied in multi-reflection time-of-flight mass spectrometry.\par
 In the scope of precision mass measurements of radioactive ions, we will focus on the availability of only two available reference masses, and later in the manuscript, even only a single reference mass at the time of the measurement. While many reference species are available in analytic mass spectrometry at the same time, for MRTOF-MS of radioisotopes, thermal ion sources -- typically alkali or alkali earths -- are generally utilized to ensure the reliable injection of single ions into the device. This sacrifice is accepted to exclude any effects caused by Coulomb interaction and to preclude detector saturation.\\%
\section{Acceleration region with a linear field increase}
\label{sec:General}
\begin{figure}[b]
\includegraphics[width=0.45\textwidth]{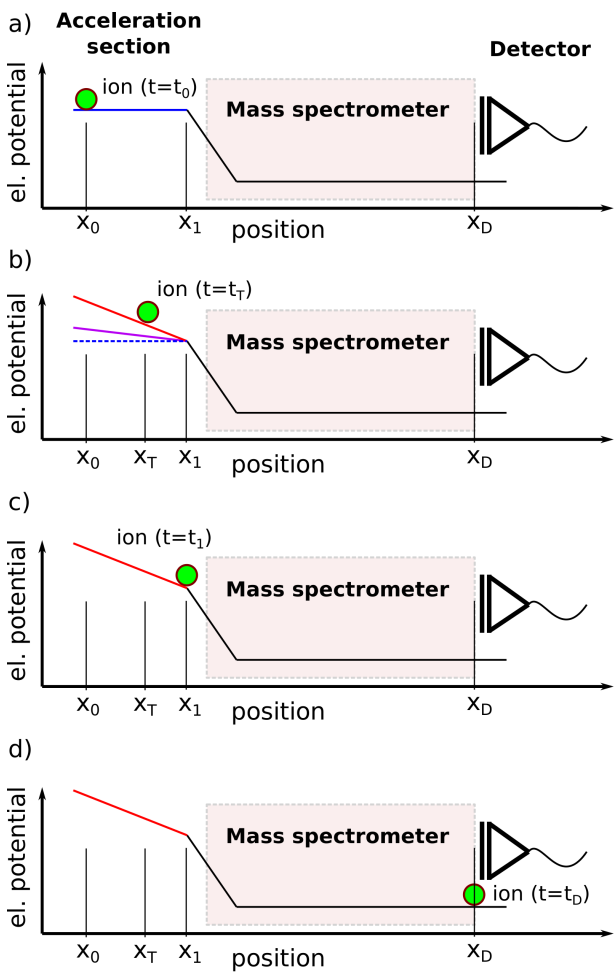}
	\caption{ \label{fig:scetch} 	Calculation sequence within the model. a) ion rests in a field-free region. b) electric ejection field starts to increase linearly in time until the full field strength is reached and the ion starts to travel towards the mass spectrometer. c) full field strength is reached and the ion travels with constant acceleration to the end of the acceleration region. d) ion travels through the mass spectrometer that is regarded as arbitrary, and reaches the detector.}
\end{figure}

We consider a simple scenario in which an ion, starting from an initially field-free region (initial trapping fields are not considered and regarded as switched off) with zero initial energy, is accelerated by an electric field increasing linear in time. In reality, switching transitions usually have a nearly exponential shape and the effects will differ from the linear model. The major intention is to cover the majority of impacts by the simplest time-dependent shape beyond the application of the Heaviside step function for the transition (as used before). On the other hand, we also note that it is possible to approach a well-shaped linear transition in the experiment by application of adapted switches and circuitries, which is studied in parallel to our present work.\\
From the time $t_0$ onwards (with $t_0 = 0$ until explicitly mentioned again), an ejection field is increasing linearly in strength with time $t$ until time $t_\mathrm{T}$, at which the transition of the switch is completed and the full field-strength $E_0$ is present. As it is not required to consider traverse motion in this basic study, we regard only one dimension where the electric field is given by:
  
\begin{eqnarray}
  \label{eq:field}
&E(t) &{}=
\begin{cases}
	E_0\cdot \frac{t}{t_\mathrm{T}} & \mathrm{for\,\,} t \leq t_\mathrm{T} \\
  E_0 & \mathrm{for\,\,} t > t_\mathrm{T} \\
\end{cases}
\end{eqnarray}
 The calculation sequence of this model is illustrated in Fig.~\ref{fig:scetch}. Starting at rest, an ion with mass $m$ and charge $q$ is accelerated and traveled the distance $x( t_\mathrm{T}) = x_\mathrm{T}$ when the field transition is complete. It further travels under the full field strength until reaching the position $x( t_\mathrm{1}) = x_1$, which is considered as the end of the acceleration region. Beyond the position $x_1$, any type of TOF mass spectrometer can follow, meaning that the fields can be considered as unknown and will be treated in a general context later. The ion travels through this spectrometer until reaching the detector at position $x_\mathrm{D}$ at the detection time $t_\mathrm{D}$.\par
In this study, the acceleration section of the device is as long or longer than the distance $x_\mathrm{T}$, so that $x( t_\mathrm{T}) \leq x_1$ (and $ t_\mathrm{T} \leq t_1$), which means that the transition is finished before the ion reaches the end of the acceleration section. Furthermore the ion starts at rest ($v_0 = v(t_0) = 0$) at position ($x_0 = x(t_0) = 0$), where $t_0 = 0$.\par
\begin{figure}[b]
\includegraphics[width=0.48\textwidth]{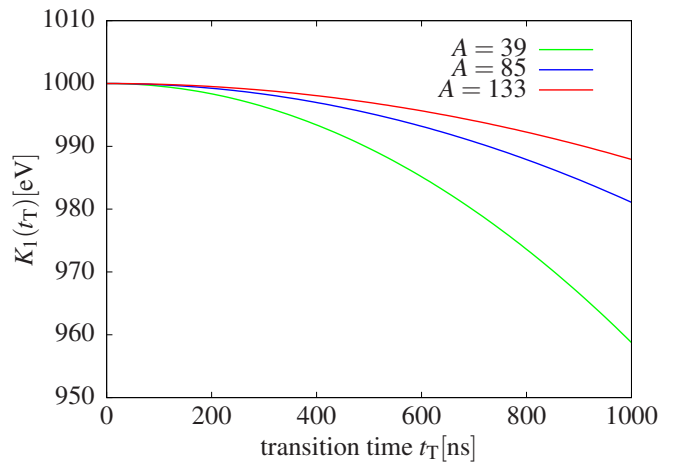}
  \caption{\label{fig:ekin} Kinetic energy obtained during ion acceleration as a function of the duration of a linear field increase. The numbers are calculated assuming $200\,\mathrm{V/cm}$ as full ejection-field strength for singly charged ions with atomic mass numbers $A=39,85,133$ traveling to the position $x_1=5\,\mathrm{cm}$ using transition times $t_\mathrm{T}$ up to $1000\,\mathrm{ns}$.}
\end{figure}

The motion of the ion as a function of time can be derived in a very straight-forward way. We can calculate the velocity $v_\mathrm{T}$ at the time $t=t_\mathrm{T}$ by:
\begin{equation}
  v_\mathrm{T} = \frac{qE_0}{m \,t_\mathrm{T}}\int_{0}^{ t_\mathrm{T}} dt \,t = \frac{qE_0}{m \,t_\mathrm{T}}\frac{t_\mathrm{T}^2}{2}\quad {,}
\end{equation}
and the corresponding position $x_\mathrm{T}$ by:
\begin{equation}
  x_\mathrm{T} = \int_{0}^{t_\mathrm{T}} dt \,v(t) = \frac{qE_0}{m}\frac{t_\mathrm{T}^2}{6}
\end{equation}
To reach the end of the acceleration section $x_1$, the ion still has to travel under the constant acceleration $\ddot{x}(t) = \frac{qE_0}{m}$ over distance of $x_1 - x_\mathrm{T}$ with initial velocity $v_\mathrm{T}$.
\begin{equation}
  x_1 =x_\mathrm{T} + v_\mathrm{T} \cdot (t_1 - t_\mathrm{T}) + \frac{qE_0}{2m} \cdot (t_1 - t_\mathrm{T})^2 
\end{equation}
The time $t_1$ which the ion needs to travel to $x_1$ is obtained by the quadratic equation:
\begin{equation}
  t_1^2 - t_\mathrm{T}\,t_1 - \frac{2 m x_1}{q E_0} + \frac{t_\mathrm{T}^2}{3} = 0
\end{equation}
with the positive solution:
\begin{equation}
	\label{t1}
  t_1 = \frac{t_\mathrm{T}}{2}+ \sqrt{\frac{2 m x_1}{q E_0} - \frac{t_\mathrm{T}^2}{12}}\quad {.}
\end{equation}
The transition time $t_\mathrm{T}$ is now explicitly present in the time-of-flight $t_1$ at that position, and the relation usually applied for mass calculations from the TOF of ions:
\begin{equation}
	\label{eq:tstat}
	t_\mathrm{D} = t_0 + \alpha \sqrt{\frac{m}{q}}\quad {,}
\end{equation}
with $t_\mathrm{D}$ being the TOF (detection time) and $\alpha$ the device constant, is not exact anymore. The kinetic energy $K(t_1) = K_1$ at the position $x_1$ is:
\begin{equation}
  K_1 = q E_0 x_1 - \frac{q^2 E_0^2 t_\mathrm{T}^2}{24m}\quad {,}
\end{equation}
which contains a mass dependency scaling with the square of the transition time $t_\mathrm{T}$. The obtained energy loss with increasing $t_\mathrm{T}$ is intuitive as the ion already travels a fraction of the distance under a decreased acceleration field before the full field is present. This leads to a reduction of kinetic energy as compared to the case at which an instant field is present. The difference of kinetic energies between the case of the static field and that of the linearly increasing field is:
\begin{equation}
	\label{delk1}
	\delta K_1 = K_1|_{t_\mathrm{T} > 0} - K_1|_{t_\mathrm{T}=0} = -\frac{q^2 E_0^2 t_\mathrm{T}^2}{24 m}\quad{.}
\end{equation}
For later usage, the static part of the kinetic energy $K_1$ for $ t_\mathrm{T} = 0$ and the dynamic contribution for $t_\mathrm{T} > 0$ are noted as 
\begin{equation}
	\label{K1}
	K_1 = q E_0 x_1 + \delta K_1 \quad {.}
\end{equation}
The effect of the field transition on $K_1$ at the position $x_1$ for increasing transition times and typical acceleration parameters is shown in Fig.~\ref{fig:ekin}. \par
\section{Modified time-of-flight for a general 1D mass spectrometer}
\label{sec:moditof}

In order to apply the dynamic ion acceleration to any mass spectrometer that follows the acceleration region, the TOF $t_\mathrm{D}$ to reach the detector can be obtained from the TOF integral starting at the end of the acceleration section $x_{1}$ and ending at the detector $x_\mathrm{D}$:
\begin{equation}
	\label{eq:tD}
	t_\mathrm{D} = t_1 + \int_{x_1}^{x_\mathrm{D}} \frac{dx}{\sqrt{\frac{2 K(x)}{m}}}\quad{,}
\end{equation}
where the kinetic energy measures the value that the ion has obtained in the acceleration section plus the integral over the charge times the electric fields $E(x)$ experienced in the mass spectrometer up to the position $x$:
\begin{equation}
	\label{Ktotal}
	K(x) = K_1 + q \int_{x_1}^{x} E(x')dx'{.}
\end{equation}
As it will turn out useful to separate the charge $q$ from the kinetic energy, this integral expression is used instead of $K(x)$ as generic term, and by substituting Eq.~\ref{K1} and Eq.~\ref{Ktotal} into Eq.~\ref{eq:tD} we obtain
\begin{equation}
	t_\mathrm{D} = t_1 + \sqrt{\frac{m}{2}}\int_{x_1}^{x_\mathrm{D}} \frac{dx}{\sqrt{q E_0 x_1 + q \int_{x_1}^{x}E(x')dx' + \delta K_1}}\quad{.}
\end{equation}
Typical transition times of modern devices are below $200\,\mathrm{ns}$ and typical average kinetic energies for TOF-MS are above $1\,\mathrm{keV}$, which causes a deficit in kinetic energy well below one per cent, if the mass number is not too low. Thus, an expansion to the first order for small $\delta K_1$ is sufficient for many realistic cases:
\begin{equation}
	\label{general}
	\begin{split}
		t_\mathrm{D} \approx t_1 + & \sqrt{\frac{m}{2}}\int_{x_1}^{x_\mathrm{D}} \frac{dx}{\sqrt{q \widetilde{K}_\mathrm{stat}(x)}}  \\
		& - \sqrt{\frac{m}{2}}\int_{x_1}^{x_\mathrm{D}} \frac{\delta K_1}{2\sqrt{q \widetilde{K}_\mathrm{stat}(x)}^3}dx \quad {,}
	\end{split}
\end{equation}
with
\begin{equation}
	\widetilde{K}_\mathrm{stat}(x) = E_0 x_1 + \int_{x_1}^{x}E(x')dx'
\end{equation}
as expression for a measure of kinetic energy (excluding the charge) at the position $x$ in the spectrometer for the static case with $ t_\mathrm{T} = 0$. While the approximation in Eq.~\ref{general} is general (as long as valid), for the linear field increase we can express the time $t_1$ and the energy deficit $\delta K_1$ using Eq.~\ref{t1} and Eq.~\ref{delk1}.
\begin{equation}
	\begin{split}
		t_\mathrm{D} &  = \frac{t_\mathrm{T}}{2}+ \sqrt{\frac{2 m x_1}{q E_0} - \frac{t_\mathrm{T}^2}{12}} + \frac{1}{\sqrt{2}} \int_{x_1}^{x_\mathrm{D}} \frac{dx}{\sqrt{\widetilde{K}_\mathrm{stat}(x)}} \cdot \sqrt{\frac{m}{q}}  \\
		& + \frac{1}{\sqrt{8}} \frac{E_0^2 t_\mathrm{T}^2}{24} \int_{x_1}^{x_\mathrm{D}} \frac{dx}{\sqrt{\widetilde{K}_\mathrm{stat}(x)}^3} \cdot \sqrt{ \frac{q}{m} }  
	\end{split}
\end{equation}
Except for the first two terms, the equation can be separated in parts where charge and mass are extracted as $\sqrt{m/q}$ and $\sqrt{q/m}$, respectively. However, as $ t_\mathrm{T}^2 /12$ is very small, the remaining square root can be expanded as
\begin{equation}
	\label{eq:squarerootexpansion}
	\sqrt{\frac{2 m x_1}{q E_0} - \frac{t_\mathrm{T}^2}{12}} \approx \sqrt{\frac{2 m x_1}{q E_0}} - \frac{1}{2}\sqrt{\frac{q E_0}{2 m x_1}} \frac{ t_\mathrm{T}^2}{12} \quad {,}
\end{equation}
where charge and mass can be extracted in the same way as mentioned before. If, for completeness, also an additional non-zero offset time for the field transition $t_0$ is added to the TOF, we can finally rewrite Eq.~\ref{general} as
\begin{equation}
	\label{eq:tdyn}
	t_\mathrm{D} = t_0 + \frac{t_\mathrm{T}}{2} + \alpha \sqrt{\frac{m}{q}} + \beta \sqrt{\frac{q}{m}} \quad {.}
\end{equation}
The new equation now explicitly contains $t_\mathrm{T}/2$, where the sum $t_0 + t_\mathrm{T}/2$ acts as a shifted (effective) offset time that can be treated the same way as $t_0$. The two parameters $\alpha$ and $\beta$ are device-specific functions depending on the electric fields in the spectrometer, but also on the initial condition of an ion (position, energy). Only if a high-quality TOF focus can be achieved at the detector, and the dependency on the initial conditions can be averaged resulting in a mean TOF of an ion distribution, $\alpha$ and $\beta$ can be regarded as true device constants. The first device constant $\alpha$ is independent from $t_\mathrm{T}$ and represents the electrostatic system without disturbance. All additional features are covered by the value of $\beta \sqrt{q/m}$, which is proportional to $ t_\mathrm{T}^2$ and reciprocal to the square root of the mass. Thus, we obtain the well-known static case from Eq.~\ref{eq:tstat} for $ t_\mathrm{T} \rightarrow 0$ and also for $ m \rightarrow \infty$.\par
We note that the approximation made in Eq.~\ref{general} does not cover the case of ion reflections as performed in a reflectron or multi-reflection device. As the ions are slowed down to zero energy, the assumption that the influence of $\delta K_1$ on time-of-flight is small enough to justify the expansion is not valid in general. However, these systems are optimized to operate in a non-dispersive (isochronous) or low-dispersive mode, \textit{i.e.} to achieve a very similar flight time at the detector for different ion energies. In that way the derivative $\partial t_\mathrm{D}/ \partial \delta K_1$ yields very small values, even if the ions are momentarily stopped.
\section{Calculation of the mass from time-of-flight data}
\label{sec:calculation_of_the_mass_from_time_of_flight_data}
\begin{figure}[t]
\includegraphics[width=0.46\textwidth]{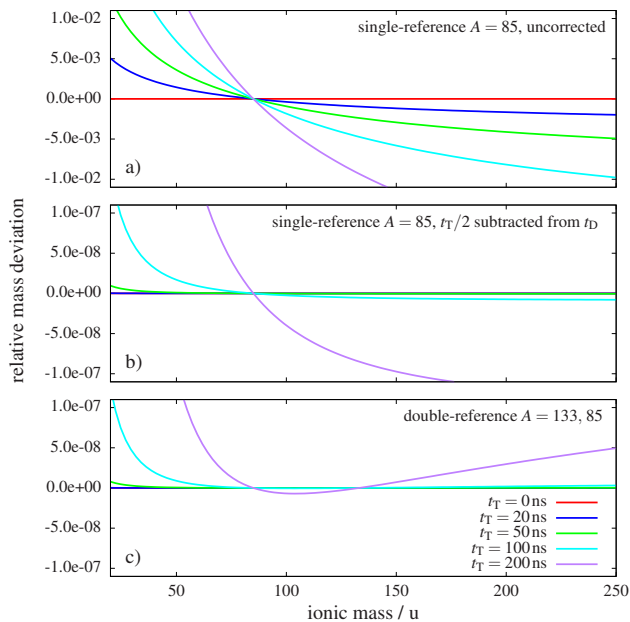}
	\caption{\label{fig:detail_mass_deviation} Relative mass deviation from the true mass as a function of the mass number in case of a finite transition time of the extraction field. The TOF is calculated for a $x_1 = 5\,\mathrm{cm}$ acceleration section ($x_0 = 0$, $t_0 = 0$) and a subsequent drift section of $10\,\mathrm{cm}$ with $1\,\mathrm{keV}$ kinetic drift energy in the case of zero transition time. Recalculated masses over a wide mass range are shown for selected transition times. a) using a single reference mass without correction for the field transition. b) using a single reference mass and subtracting the constant offset term $ t_\mathrm{T}/2$ (Eq.~\ref{eq:tdyn}) from the TOF. c) using two reference masses.}
\end{figure}

Reviewing first the static case, masses can be calculated from the time-of-flight using a single reference mass $ m_1$, two reference masses $ m_1, m_2$, or even several masses if available during the measurement. We will review the calibration with a single reference mass and two reference masses. When calibrating with a single reference species with well known mass ($m_1$ with the TOF $t_\mathrm{D1}$), the device parameter $\alpha$ from Eq.~\ref{eq:tstat} can be eliminated by division $ t_\mathrm{D}/ t_\mathrm{D1}$ ($ t_\mathrm{D}$ for the analyte ion), and the solution for the mass of interest is:
\begin{equation}
  \label{eq:single}
  m = m_1\cdot \left(\frac{t_\mathrm{D} - t_0}{t_\mathrm{D1} - t_0}\right)^2 \quad {.}
\end{equation}
The offset time $t_0$, denoting the real starting time of the ejection process as compared to the start time of the data acquisition system in the experiment, is unknown and must be determined with other methods. To overcome this problem, $t_0$ can either be measured \cite[]{SCHURY201419} or can be eliminated from the mass equation using the information of a second mass if available (see, {\emph e.g.}, \cite[]{Wienholtz2013}), referred to as double-reference calibration:
\begin{equation}
  \label{eq:double}
  m = \left(C_\mathrm{TOF}\cdot \Delta_{1,2} + \frac{\Sigma_{1,2}}{2}\right)^2
\end{equation}
where $\Delta_{1,2} = \sqrt{m_1} - \sqrt{m_2}$, $\Sigma_{1,2} = \sqrt{m_1} + \sqrt{m_2}$, and
\begin{equation}
  C_\mathrm{TOF} = \frac{ \left( 2 t_\mathrm{D} - t_\mathrm{D1} - t_\mathrm{D2} \right) }{ 2\left(t_\mathrm{D1} - t_\mathrm{D2}\right) }
\end{equation}
A mass calibration with the linear model for $ t_\mathrm{T} > 0$ obtains the difference that the additional unknown coefficient $\beta$ is present. Three different masses are now required to extract all unknown coefficients and are obtained from the linear system:
\begin{eqnarray}
	\label{linear-system}
\begin{cases}
t_\mathrm{D1} = t_0 + \frac{t_\mathrm{T}}{2} + \alpha X_{1} + \frac{\beta}{X_{1}} & \\
t_\mathrm{D2} = t_0 + \frac{t_\mathrm{T}}{2} + \alpha X_{2} + \frac{\beta}{X_{2}} & \\
t_\mathrm{D3} = t_0 + \frac{t_\mathrm{T}}{2} + \alpha X_{3} + \frac{\beta}{X_{3}} & \quad {.}
\end{cases}
\end{eqnarray}
\begin{widetext}
This can be solved in a straight forward way, yielding:
\begin{eqnarray}
	\label{three-mass-calibration}
t_0 + \frac{t_\mathrm{T}}{2} &=& \frac{t_\mathrm{D1}X_{1}({X_{2}}^{2}-{X_{3}}^{2})+t_\mathrm{D2}X_{2}({X_{3}}^{2}-{X_{1}}^{2})+t_\mathrm{D3}X_{3}({X_{1}}^{2}-{X_{2}}^{2})}{(X_{1}-X_{2})(X_{2}-X_{3})(X_{3}-X_{1})} \\
\alpha &=& \frac{t_\mathrm{D1}X_{1}({X_{3}}-{X_{2}})+t_\mathrm{D2}X_{2}({X_{1}}-{X_{3}})+t_\mathrm{D3}X_{3}({X_{2}}-{X_{1}})}{(X_{1}-X_{2})(X_{2}-X_{3})(X_{3}-X_{1})} \\
	\beta &=& \frac{X_{1}X_{2}X_{3}\{t_\mathrm{D1}({X_{3}}-{X_{2}})+t_\mathrm{D2}({X_{1}}-{X_{3}})+t_\mathrm{D3}({X_{2}}-{X_{1}})\}}{(X_{1}-X_{2})(X_{2}-X_{3})(X_{3}-X_{1})} \quad {,}
\end{eqnarray}
\end{widetext}
with masses and charges $X_i = \sqrt{m_i/q_i}$. As we want to focus additionally on the case of highly stable ion sources with very limited reference masses, the problem can be reduced if we assume that the electronic components defining the shifted offset time $t_0 + t_\mathrm{T}/2$, once accurately determined, do not change significantly. Two well-known reference masses are then required and their corresponding times-of-flight. Starting from Eq.~\ref{eq:tdyn} with $t_0 + t_\mathrm{T}/2$ determined before, the constants $\alpha$ and $\beta$ can be found by using only two reference ions with mass and charge $X_1 = \sqrt{m_1/q_1}$ and $X_2 = \sqrt{m_2/q_2}$, and detection times $ t_\mathrm{D1}$ and $ t_\mathrm{D2}$, respectively:
\begin{equation}
	\label{eq:Afunc}
	\alpha = \frac{\widetilde{t}_\mathrm{D1} X_1 - \widetilde{t}_\mathrm{D2} X_2}{X_1^2 - X_2^2}
\end{equation}
and
\begin{equation}
	\label{eq:Bfunc}
	\beta = \frac{\widetilde{t}_\mathrm{D2} X_1^2 X_2 - \widetilde{t}_\mathrm{D1} X_1 X_2^2}{X_1^2 - X_2^2} \quad {.}
\end{equation}
 To calculate the mass of an unknown ion, the detection time can be reduced by the shifted offset time $t_0 + t_\mathrm{T}/2$
\begin{equation}
	\widetilde{t}_\mathrm{D} = t_\mathrm{D} - \frac{t_\mathrm{T}}{2} - t_0 \quad {,}
\end{equation}
and the equation for the reduced detection time of an unknown ion with $X = \sqrt{m/q}$ is
\begin{equation}
	\widetilde{t}_\mathrm{D} = \alpha\cdot X + \frac{\beta}{X} \quad {,}
\end{equation}
giving a quadratic equation for the mass-to-charge ratio
\begin{equation}
	X^2 - \frac{\widetilde{t}_\mathrm{D} X}{\alpha} + \frac{\beta}{\alpha} = 0 \quad
\end{equation}
with the positive solution
\begin{equation}
	X = \frac{\widetilde{t}_\mathrm{D}}{2\alpha} + \sqrt{ \left(\frac{\widetilde{t}_\mathrm{D}}{2\alpha}\right)^2 - \frac{\beta}{\alpha}} \quad {.}
\end{equation}
Finally, the mass is obtained by
\begin{equation}
	\label{eq:finalmass}
	m = q\left(\frac{\widetilde{t}_\mathrm{D}}{2\alpha} + \sqrt{ \left(\frac{\widetilde{t}_\mathrm{D}}{2\alpha}\right)^2 - \frac{\beta}{\alpha}}\right)^2 \quad {.}
\end{equation}
For illustration of the deviation of the calculated mass from the true mass value when the raising time of the extraction field is not included, a simple example for a single-stage TOF spectrometer has been calculated. Positively charged ions starting at $x_0 = 0$ for $t_0 = 0$ are accelerated in an ideal dipole field of $200\,\mathrm{V/cm}$ strength up to the position of $x_1 = 5\,\mathrm{cm}$. After the travel through this distance, the ions move without any force through a drift tube of length $L = x_D - x_1$. The first-order TOF focus in the electrostatic case will occur using $L = 2(x_1 - x_0) = 10\,\mathrm{cm}$ (see \cite[]{Wiley1955}) and a TOF detector is assumed at this position. In order to include the raising time of the field, Eq.~\ref{eq:tD} has been used, where the integral on the right side becomes trivial due to the constant velocity. The time-of-flight for a mass $A=100$ ion of this system yields about $4.6\,\mathrm{\mu s}$. The determination of the mass from the TOF has then been performed with Eq.~\ref{eq:single} and Eq.~\ref{eq:double} to investigate the relative difference of the measured mass from the real mass $ (m_\mathrm{meas} - m_\mathrm{real}) / m_\mathrm{real}$. The mass of an $A=85$ ion (close to the alkali $^{85}$Rb) will be considered as reference mass for calibration using Eq.~\ref{eq:single}, and the mass $A=133$ (like $^{133}$Cs) is additionally used as a second reference when Eq.~\ref{eq:double} is applied for two reference ions. The result for both methods of mass calibration is shown in Figure~\ref{fig:detail_mass_deviation}.\par
In the case where a single reference mass is used and no further corrections are done as in Fig.~\ref{fig:detail_mass_deviation} a), deviations in the order of a few parts per thousand up to a few per cent are obtained from this example, where the latter could result in a mismatch of one mass number or even more. It is important to note that the transition time in this number example is already a large fraction of the total time-of-flight (see Sec.~\ref{sec:t0influence} for more details). In Fig.~\ref{fig:detail_mass_deviation} b), the constant term $ t_\mathrm{T}/2$ from Eq.~\ref{eq:tdyn} has been subtracted and the residual deviations are orders of magnitude lower, which proves its large contribution to the inaccuracies. It can be further seen that the masses cross the zero line at the calibration mass as expected. The calibration method with two masses in Fig~\ref{fig:detail_mass_deviation} c) performs very well between the two reference points and for usual operation the mass deviations can in principle be regarded as negligible, even though the effect of finite transition times does not fully vanish. A full correction requires three masses as in Eq.~\ref{linear-system}.\\
\section{Numerical studies}
\label{sec:numerical_studies}
\begin{figure*}[ht]
	\centering
	\includegraphics[width=0.7\linewidth]{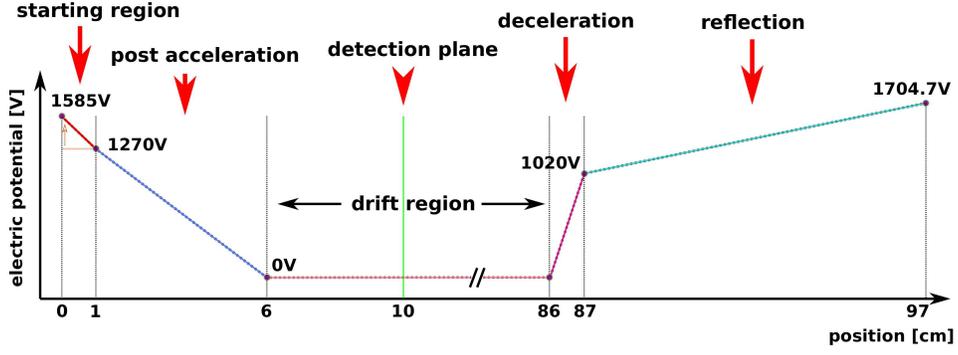}
	\caption{Electric potential as a function of the position for the chosen example of an ideal reflectron in one dimension. Ions are initialized at the position $x_0 = 0.5\,\mathrm{cm}$ and after the reflection, on the way back to the starting region, the ion detection takes place at the position $x = 10\,\mathrm{cm}$.}
	\label{fig:reflectron}
\end{figure*}
In the following, a finite transition time will be studied by calculating the ions' time-of-flight within a model of a classical reflectron \cite[]{Mamyrin1973} in one dimension. The device characterizing functions $\alpha$ and $\beta$ will be discussed for single ions, and the masses of ions calculated using a simulation with many ions yielding an average time-of-flight. An additional constant offset time of the measurement will not be considered ($t_0 = 0$). \par
The reason for the choice of a reflectron is that we cover the situation where a reflection of ions takes place and hence probe the robustness of the approximations in Eq.~\ref{general} for such scenarios. The reflectron consists of five discrete separated regions, where in each of them a constant electric field of certain strength is present. Although such a system can be calculated in an analytical way, a numerical solution has been chosen to ensure more flexibility, such as adding/removing field regions during the optimization. A description of the reflectron model including the applied electric potentials is shown in Fig.~\ref{fig:reflectron}.
The field transition between each two regions is instantaneous, which is an approximation referring to the usage of conductive meshes or wire grids. The ions are accelerated using two different dipole fields, which is a linearly increasing field (in time) in the starting region followed by a constant field for post acceleration. They pass through a field-free drift region of $80\,\mathrm{cm}$ length, and are later decelerated using two other regions with dipole fields directed against die ions' motion: a strong field over a short distance followed by a weak field over a longer distance, until a reflection is achieved. After the reflection, the ions return to the origin and are focused at the detection plane position of $x = 10\,\mathrm{cm}$ where their times-of-flight (or also relative positions) and velocities can be recorded.\par
\begin{figure}[t]
	\centering
	\includegraphics[width=0.9\linewidth]{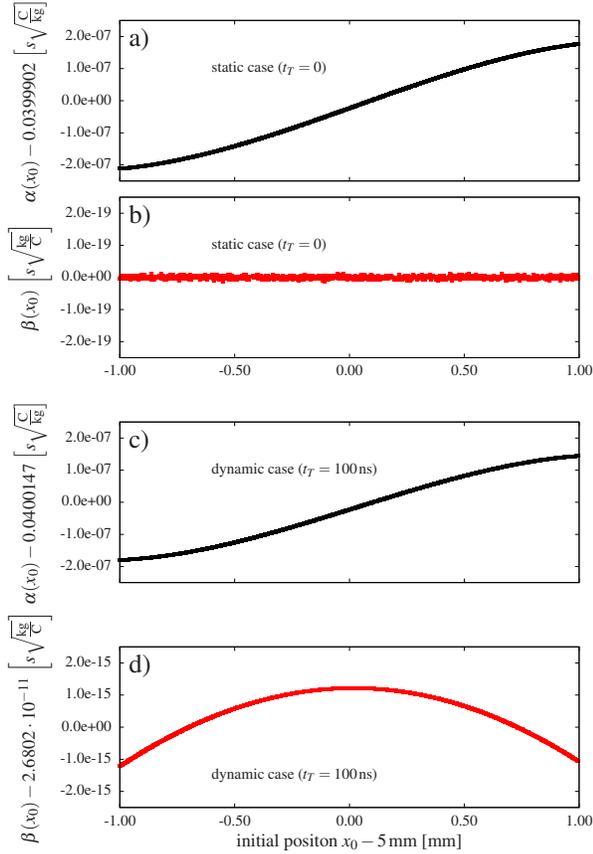}
	\caption{$\alpha$ and $\beta$ as a function of the initial position $x_0$ of the ions at the chosen focal point. a) and b) static field without transition time of the ejection field. c) and d) transition time of $t_\mathrm{T} = 100\,\mathrm{ns}$.}
	\label{fig:Device-functions}
\end{figure}

The characteristic device functions $\alpha(x_0)$ and $\beta(x_0)$ written as constants in Eq.~\ref{eq:tdyn} describe the time-of-flight as a function of the ions' mass and charge, but depend on the initial conditions of the ions. In the simple case of zero initial velocity as regarded in this study, the initial position of the ion is the only parameter. Note that all voltages and also the position of the detection plane modify the entire function and are regarded as fixed. To extract the device functions using Eq.~\ref{eq:Afunc} and Eq.~\ref{eq:Bfunc}, the time-of-flight of two single ions have been simulated for different initial positions. We assume two ion species of integer mass values, in this case $85.0$ and $133.0$, both assumed as positive and singly charged ions (close to typical alkali masses). The result of the device functions for the voltages and focus position discussed below is shown in Fig.~\ref{fig:Device-functions}.
 For the static case shown in Fig.~\ref{fig:Device-functions} a) and b), the function $\alpha(x_0)$ has an offset which corresponds to the main contribution to the device constant $\alpha$ and a variation according to the focal conditions (achieved by voltage tuning as discussed below), which in the present case is five orders of magnitude smaller than the offset (see y-axis). As expected from the discussions in Sec.~\ref{sec:moditof}, the function $\beta(x_0)$ results in values equal to zero within the precision of the simulation. If a transition time of $t_\mathrm{T} = 100\,\mathrm{ns}$ is included as in c) and d), the function $\alpha(x_0)$ changes slightly and now also non-zero values for $\beta(x_0)$ are obtained. Note that value of $\beta(x_0)$ is to be multiplied by the reciprocal $\sqrt{q/m}$ being about six orders of magnitude larger than $\sqrt{m/q}$ for a singly charged ion with $A=100$. In a mass measurement using TOF spectra by summing many ions from a spatial distribution into a histogram, the average result for $\alpha(x_0)$ and $\beta(x_0)$ from the entire initial distribution of $x_0$ values reduces these functions to the coefficients in Eq.~\ref{eq:tdyn}, which link the mass and the charge to the resulting time-of-flight.\par
\begin{figure}[h!]
	\centering
	\includegraphics[width=0.9\linewidth]{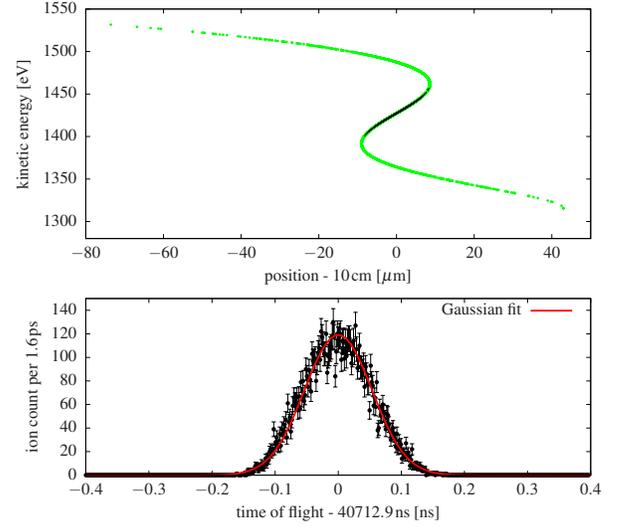}
	\caption{Top: Kinetic energy of the ions at the focus point as a function of the position at the moment when the test ion crosses the detection plane. The green points are from an initial spatial Gaussian distribution using $10000\,\mathrm{ions}$ with a width parameter of $1\,\mathrm{mm}$ in the starting region, and the black points show another $2000\,\mathrm{ions}$ with a width parameter of $250\,\mathrm{\mu m}$ mapping only the approximately linear part of the position-energy distribution at the focal point (for details, see text). Bottom: Example TOF spectrum with Gaussian fit of the time-of-flight distribution using ions initialized similar to those being the black points in the top part.}
	\label{fig:focus-phase-space}
\end{figure}
 For the simulations of time-of-flight spectra ions have been initialized using a Gaussian probability distribution around the position $x_0 = 0.5\,\mathrm{cm}$ (\textit{i.e.} the center of the starting region). In order to find acceptable focal conditions with a good overview over the position-energy distribution, an initial width parameter of $1\,\mathrm{mm}$ has been chosen. One of the ions is initialized exactly at the center position and serves as the test ion. If the test ion crosses the detection plane on the way back from the reflection in the ion mirror, the positions and energies for all other ions are recorded at that time and hence the position-energy distribution can be analyzed. The voltages have been chosen to match realistic conditions, so that comparable values can be found in a laboratory. The presence of two deceleration regions allows focal corrections of second order (of the time-of-flight as a function of the ions' starting position). A satisfying focus has been found by varying the rightmost two voltages and investigating the kinetic energy as a function of the ions' position around the focus point. The focal distribution used for further calculations is shown in Fig.~\ref{fig:focus-phase-space}.\par
The s-shape of the curve shows mainly the uncorrected third-order components of the position-energy distribution remaining after the optimization. For the data-analysis of TOF spectra, it was necessary to limit the spatial region for the initialization of ions. Although the full shape could be analyzed in principle, the representation in time-of-flight is challenging as it differs by far from a well-known shape such as a Gaussian distribution. To this end, the linear region of the position-energy distribution at the focal point is used to map the initial Gaussian distribution (width parameter of $250\,\mathrm{\mu m}$) in space approximately to a final Gaussian distribution in TOF (see the bottom of Fig.~\ref{fig:focus-phase-space}). This linear region contains an energy spread of about $50\,\mathrm{eV}$ obtained by the starting position of the ions.\par
In order to investigate the accuracy of the calculated mass corrections with a finite field-transition time, mass calculations with relative accuracies of $\delta m / m \approx 10^{-7}$ are desired. For the calculation of very precise times-of-flight, a separate code has been written. As the equations of motion do not solve for the time-of-flight but for the position of the particle, the uncertainty for the TOF is on the order of the time step itself. In the cases discussed here, this would require a time step of about $10^{-14}\,\mathrm{s}$ to reach the required precision, even though the propagation in the regions of constant fields is trivial. One way to solve for the time-of-flight more effectively while keeping the simulation flexible, and without rewriting the equations of motion to solve for the time, is the use adaptive time steps when an ion is approaching a chosen position where the electric field changes (and also the final plane for detection). When an ion crosses a plane, the time step is turned back and the crossing procedure is repeated with half time steps. The step is reduced further by such bisection down to $10^{-20}\,\mathrm{s}$. After crossing the test plane at minimum time step, it is increased back to a default value.\par
The times-of-flight have been fitted using a Gaussian with the binned maximum-likelihood estimator provided by the ROOT package \cite[]{BRUN199781} developed at CERN. From the extracted center of the distribution the characteristic device functions, the masses, and the relative mass uncertainties have been calculated (on the order of $10^{-7}$ or below if 50000 ions are used for one spectrum). The study has been performed once for the static approach, \textit{i.e} without transition time of ion extraction, and once for $t_\mathrm{T} = 100\,\mathrm{ns}$.\par
\begin{figure}[t]
	\centering
	\includegraphics[width=0.8\linewidth]{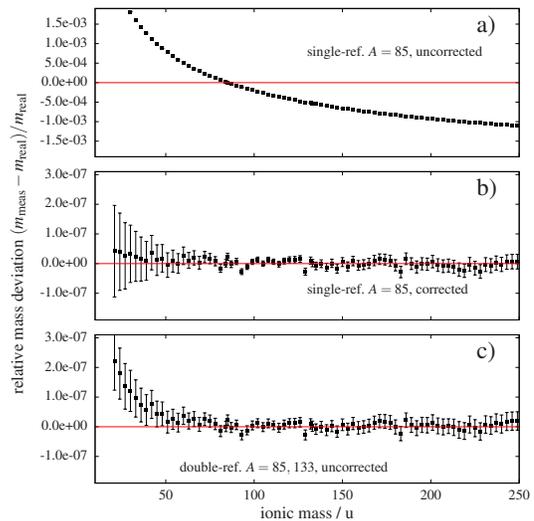}
	\caption{Relative deviation from the real mass value (upon recalculation of the mass from the TOF data of the reflectron simulation) as a function of the mass number, using a transition time of $100\,\mathrm{ns}$. a) Recalculation of masses using the static approach from Eq.~\ref{eq:single} with mass $A=85$ as a reference. b) Recalculation with correction using Eqs.~\ref{eq:finalmass}, \ref{eq:Afunc}, and \ref{eq:Bfunc}. c) Recalculation using the static approach with two reference masses from Eq.~\ref{eq:double}.}
	\label{fig:mass-result}
\end{figure}

As a next step, the relative mass accuracy for the corrections made in Sec.~\ref{sec:General} -- \ref{sec:calculation_of_the_mass_from_time_of_flight_data} has been calculated using a Gaussian fit to the data for a set of 50000 ions, where the time-of-flight of ions with mass numbers between $A=20$ and $A=250$ has been simulated using a transition time of $t_T = 100\,\mathrm{ns}$. The calculation of the mass from the obtained data has been performed using TOF spectra of two ion species, again with mass number $A=85$ and $A=133$, for the calculation of the device constants $\alpha$ and $\beta$ (regarded as a pre-calibration before the experiment). Once the device constants have been determined, the ions with $A=85$ have further been used as the single reference mass to provide the time $\widetilde{t}_\mathrm{D}$ in Eq.~\ref{eq:finalmass} for the calculation of masses of all other species (having integer mass numbers). For comparison also the calibration using Eq.~\ref{eq:single} and Eq.~\ref{eq:double} has been performed, \textit{i.e.} without correcting for the transition time, where both chosen reference masses $A=85,133$ have been used for the two required simultaneously measured references in the latter case. The result is shown in Fig.~\ref{fig:mass-result}.\par
The results approximately reflect the numbers obtained for Fig.~\ref{fig:detail_mass_deviation}, except for the new case including the full correction. Regarding first the uncorrected case in Fig.~\ref{fig:mass-result} a) with the single reference mass $A=85$, the deviation of the calculated mass values increases to the per cent level for light masses. When reference mass and analyte mass are very close to each other, such as in case of isobars, the mass deviation becomes very small and the effect of the transition time can be neglected. In Fig.~\ref{fig:mass-result} b) the full correction within the model studied has been applied by first calculating the two coefficients $\alpha$ and $\beta$ using the times-of-flight for $A=85$ and $A=133$, and subsequently the mass according to Eq.~\ref{eq:finalmass} using $A=85$ as reference mass. The relative accuracy achieved is below $5\cdot 10^{-8}$ for the full investigated mass scale. The small but significant deviations of some of the points with smallest error bars are believed to be caused by randomization of initial conditions and by possible biases that can occur by deviations of the final TOF distribution from a Gaussian shape. The slight increase of the mass deviation towards light ions reflects the accuracy limit of the approximations made in Eq.~\ref{general} (magnitude of $\delta K_1$) and Eq.~\ref{eq:squarerootexpansion} (mass value), which provide better accuracy for heavier masses. However, also the uncertainties derived from the derivatives (Sec.~\ref{sub:uncertainty-calculation_simple_case}) increase strongly for lighter masses. The usage of a two-reference calibration according to Eq.~\ref{eq:double} in Fig.~\ref{fig:mass-result} c) shows satisfying results as well, as the largest contribution comes from the additional offset time, which is eliminated for double referencing. The non-trivial contributions show significance and increase above $10^{-7}$ only for very light masses compared to the reference masses.\par
The uncertainties of the mass values have been calculated with the assumption that the mass values of the references are exactly known. Each case in Fig.~\ref{fig:mass-result} has been treated separately using the corresponding derivatives, while we will only provide the new derivatives for the corrected case in Fig.~\ref{fig:mass-result} b) in the appendix.\par
\section{A time rescaling technique for mass measurements of radioactive ions with a single-reference on-line calibration}
\label{sec:precision_measurements}
When performing precision MRTOF mass measurements of unknown exotic nuclei produced at on-line radioactive ion beam facilities, the precision and accuracy of time-of-flight measurements is of major importance. Ionic masses of nuclei of interest are measured with high accuracy and precision to provide new mass data of nuclides far from stability. A continuous mass calibration with well-known reference ion species becomes essential during the measurement and it is not guaranteed that such a species is present within the ions delivered by the facility. In facilities using the ISOL technique (see, \textit{e.g.} \cite[]{RAVN1979201,VILLARI2001465}), typically several ion species with similar mass-to-charge ratio are produced and delivered simultaneously, which provides ease for the calibration of the mass spectrometer as one or several species in the ensemble will often have well-known masses. In other experiments, such as those performed with in-flight fission and fragmentation utilizing in-flight particle separators (see, \textit{e.g.} \cite[]{Harss2001,YANO20071009}), this is not always the case and a separate ion source has to be used. Although ion-sources providing molecular ions at many mass numbers are available in principle, thus far it has been typical to use thermal ion sources delivering low-intensity beams of alkali or alkali earth ions. The primary reasons for this are two-fold. Sources of molecular ions, such as from electrospray (see \cite[]{NAIMI201324}), produce ions across a wide range of mass-to-charge ratios, necessitating extensive analysis of the reference spectrum to identify the various components; the source material also has to be replenished regularly. The thermal ion sources, however, are typically able to provide easily reproducible intensities of simple-to-identify ions enabling to reliably inject single ions into the spectrometer, and are without need of maintenance over years of operation. While a calibration with other ion sources or source conditions providing two or more ion species can be done before the experiment, a reliable calibration method is continuously required when the on-line experiment is carried out.\par
\begin{figure}[t]
	\centering
	\includegraphics[width=0.90\linewidth]{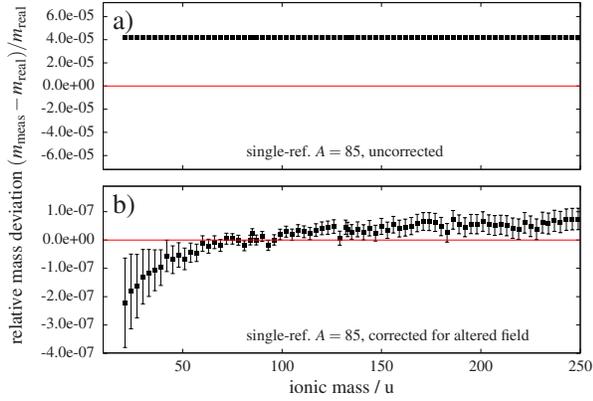}
	\caption{Relative deviation of the calculated mass value as a function of the real mass value (as in Fig.~\ref{fig:mass-result}) for a transition time of $100\,\mathrm{ns}$. Ions for calculation of $\alpha$ and $\beta$ are simulated with other conditions of the electric field as compared to all other ions including the reference at $A=85$ (repeated) at a "later" state of the setup. a) Recalculation of masses using the corrected approach from Eq.~\ref{eq:finalmass} which includes $t_\mathrm{T} > 0$, but without correction of field drifts after the first calibration. b) Recalculation using the approach for $t_\mathrm{T} > 0$ and further correcting the TOF using Eq.~\ref{eq:online_recalib}.}
	\label{fig:online-recalibration}
\end{figure}

We have investigated the feasibility of measuring the device constants $\alpha$ and $\beta$ in an independent calibration measurement and using only one reference ion species in an experiment at a later time (on-line experiment) to correct for time-of-flight drifts caused by time-dependent variations in the electric fields and thermal expansions of the spectrometer. To this end the simulation performed in Sec.~\ref{sec:numerical_studies} has been repeated for the two masses $A=85$ and $A=133$ to obtain the times-of-flight $ t_\mathrm{D1}$ and $ t_\mathrm{D2}$ for the given electric field. Afterwards, the electric field strength in the reflection section (see Fig.~\ref{fig:reflectron}) has been multiplied by a factor of $1+10^{-4}$ to simulate a later state of the setup at which the power supplies for the electrodes yield slightly different voltages. The device constants from the first simulation have been used to calculate the masses as done previously in Sec.~\ref{sec:numerical_studies}, and the reference ion with $A=85$ of the second simulation with altered field, yielding $\widetilde{t}^{\prime}_\mathrm{D1}$ (all time values from second run marked with prime), was used to scale all times-of-flight from other ions according to
\begin{equation}
	\label{eq:online_recalib}
	^{ \mathrm{corrected}}\widetilde{t}^{\prime}_\mathrm{D} = \widetilde{t}_\mathrm{D}^{\prime} \frac{\widetilde{t}_\mathrm{D1}}{\widetilde{t}_\mathrm{D1}^{\prime}} \quad{,}
\end{equation}
to match the presently measured times with the previously measured times for $\alpha$ and $\beta$, which are not newly determined during the experiment. Also in this study we assume that the start time of the ion ejection and the transition time are already measured with sufficient accuracy. The extended equation for the mass in this scenario is given by
\begin{equation}
	\label{eq:finalmass-extended}
	m = q\left(\frac{\widetilde{t}_\mathrm{D}^{\prime}}{2\alpha} \frac{ \widetilde{t}_\mathrm{D1} }{\widetilde{t}_\mathrm{D1}^{\prime}}  + \sqrt{ \left(\frac{ \widetilde{t}_\mathrm{D}^{\prime} }{2\alpha}\frac{ \widetilde{t}_\mathrm{D1} }{ \widetilde{t}_\mathrm{D1}^{\prime} }\right)^2 - \frac{\beta}{\alpha}}\right)^2 \quad {,}
\end{equation}
which is the same equation as Eq.~\ref{eq:finalmass}, but with a rescaled time for the reference mass during the measurement. We can now compare the relative mass deviation obtained in the experiment when such a recalibration is used, see Fig.~\ref{fig:online-recalibration} a), or not used, see Fig.~\ref{fig:online-recalibration} b).
The deviation without compensating for the altered electric field acts as an offset for the mass scale (Fig.~\ref{fig:online-recalibration} a), and yields about $4\cdot 10^{-5}$ for our reflectron example. This is expected as all flight times are equally scaled by the altered field. In that way, a rescaling of all flight times according to the change measured for the reference mass provides a correct mass also for any other ion.. However, the non-trivial contribution of the recalibration can be seen in Fig.~\ref{fig:online-recalibration} b). Towards masses significantly lighter than the reference mass, deviations larger than the $10^{-7}$ level are obtained due to the reuse of $\alpha$ and $\beta$ without modification.\par
From our example, we obtain the information that an on-line recalibration using a simple rescaling is meaningful also when the finite transition for ion ejection is present. Hence, a single-reference measurement can provide accurate mass results if the measured mass is not excessively far from that of the reference ions. Especially for MRTOF mass measurements, such a calibration is meaningful, as the number of ion reflections can be varied during an experiment. The newly obtained time-of-flight can then be adapted to the originally measured device constants $\alpha$ and $\beta$ by scaling. As in some cases indeed only one reliably produced reference species is available during the on-line measurement, this recalibration method is of practical use. Dependent on the real length of the transition time, special care may become important for the mass deviations towards heavy ions, which could apply for mass measurements of heavy isotopes and superheavy elements calibrated with lighter alkali ions.\par
\section{Influence of an offset time mismatch on MRTOF mass measurements}
\label{sec:t0influence}
\begin{figure}[]
	\centering
	\includegraphics[width=0.90\linewidth]{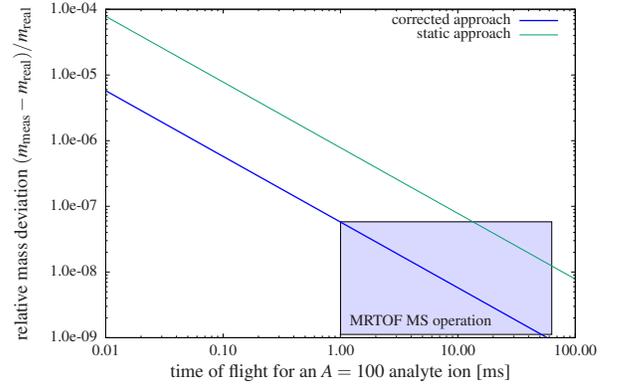}
	\caption{Relative deviation of the calculated mass value from the initial mass as a function of the total flight time for an analyte ion of mass $A=100$ with a mismatch of $5\,\mathrm{ns}$ for the offset time of the measurement. The results for the static approach with a single reference $A=85$ (green line) and the corrected approach using two references $A=85,133$ for calculation of $\alpha$ and $\beta$ and the single reference $A=85$ to calculate the mass (blue line) are compared directly by setting $t_\mathrm{T} = 0$. The rectangle shows the typical TOF region of MRTOF-MS operation. Note the logarithmic scale of both axes.}
	\label{fig:t0-mass-deviation}
\end{figure}
We want to revisit a particular case of calibration, which has been introduced at an MRTOF setup at the RIBF facility of RIKEN in Japan. For the operation of this setup, the presence of more than one available reference mass is not always guaranteed during the measurement and a different measurement method for $t_0$ has been introduced \cite[]{SCHURY201439} using electronic signals from the switch. This way of calibration carries some intrinsic uncertainties as also discussed in \cite[]{SCHURY201439} and the duration of the field transition was not included in the calibration. However, the measured masses turned out to be correct and a prominent effect of the field transition could not be recognized. In the scope of this work, we want to revisit the possible mismatch of the electric response versus the time $t_0 + t_\mathrm{T}/2 = \widetilde{t}_0$ and its influence in case of long times-of-flight. In case of a finite transition time, the sum $t_0 + t_\mathrm{T}/2 = \widetilde{t}_0$ can be regarded as a single source of uncertainty as both of them add up directly. From a theoretical point of view, the mismatch between the real offset time $t_0$ (before the voltage is switched) and the effective offset time $\widetilde{t}_0$ is then $t_\mathrm{T}/2$. It is worthwhile to look at experimental parameters including those for typical MRTOF-MS operation. We will compare the situation when the offset time was determined by signal measurements, assuming a mismatch of $\widetilde{t}_0$, with a case when more than one reference mass was available the corrected approach of Eq.~\ref{eq:finalmass} could be used with the same mismatch.\par
For the latter case, we assume again two reference measurements of the same measurement set (other than in Sec.~\ref{sec:precision_measurements}) to determine $\alpha$, $\beta$, and the mass $m$ of an $A=100$ ion regarded as an unknown analyte. The relative mass deviation with a constant mismatch $\Delta \widetilde{t}_0 = 5\,\mathrm{ns}$ has been investigated for different total flight times.\par
As it has been demonstrated that the full correction for $t_\mathrm{T}>0$ (see Fig.~\ref{fig:mass-result}) yields accurate results, and to be able to compare with the static approach from Eq.~\ref{eq:single} in a simple way, we only consider a static system with a mismatch of the offset time. With this simplification the correct flight times can then directly be calculated by scaling the TOF obtained for an $A=100$ ion and no further separate simulation is necessary to obtain longer TOF values. The offset time mismatch is added and $\alpha$, $\beta$, and the masses are recalculated yielding a deviation of calculated mass from the real value as shown in Fig.~\ref{fig:t0-mass-deviation}. For typical flight times of several milliseconds and the chosen analyte mass and references, both scenarios provide satisfying accuracies which can explain the non-recognition of the transition time effect. As two masses are used for the corrected approach (in this static case quite similar to the double-reference calibration from Eq.~\ref{eq:double}), the relative deviations drop quickly below the $10^{-7}$ level, which approaches the accuracy limits of state-of-the-art devices. The corrected method allows to perform a single-reference measurement at the moment of the experiment using a calibration as discussed in Sec.~\ref{sec:precision_measurements}, but converges faster as it is technically a double-reference method.
\section{Summary and Conclusion}
\label{sec:conclusion}
A new time-of-flight model including a linear increase in field strength for the ejection of ions towards a TOF mass spectrometer has been investigated. Using first-order approximations, a general form for the flight time of ions in this scenario has been derived resulting in the opportunity to correct for the emerging distortions of calculated masses from TOF data. This correction covers all usual time-of-flight mass spectrometers and also multi-reflection devices. The magnitude of the effects has been investigated using dedicated numerical examples.\par
From a theoretical point of view, the results show effects being of significance for single-reference mass measurements as it can lead to a misidentification in the order of a mass number if the trap-ejection switches are not state-of-the-art. The corrections could also play a general role for high-precision mass measurements with MRTOF-MS when the relative mass precision is well below $\delta m / m = 10^{-7}$. On the other hand, as the largest contribution of impact is an additional constant time offset of half the duration of the ejection-field transition, the influence of the effect turns out to be negligibly small if a calibration of the measurement offset time with two appropriate reference masses is possible.\par
The field-transition times have not been recognized as remarkable error source in practice. In case of MRTOF-MS, the total flight times are on the order of milliseconds and in many cases ions of a similar mass to that of the analyte can be used as a reference. Furthermore the transition times of state-of-the-art switches are very short, \textit{i.e.} on the order of tens of nanoseconds.\par
Test measurements for confirmation of this study are to be performed in an upcoming experiment under realistic conditions, where also the robustness of the approach for deviations of the transition from a linear shape can be investigated. So far, conclusive experimental data is not present and requires measurements including a larger mass region and a dedicated setup with adjustable ejection field transition.\\\par 
\section{Acknowledgements}
\label{sec:acknowledgements}
This work was supported by the Japan Society for the Promotion of Science KAKENHI (Grants No. 2200823, No. 24224008, No. 24740142, No. 15H02096, and No. 15K05116, 17H01081).
%



\begin{thebibliography}{29}%
\makeatletter
\providecommand \@ifxundefined [1]{%
 \@ifx{#1\undefined}
}%
\providecommand \@ifnum [1]{%
 \ifnum #1\expandafter \@firstoftwo
 \else \expandafter \@secondoftwo
 \fi
}%
\providecommand \@ifx [1]{%
 \ifx #1\expandafter \@firstoftwo
 \else \expandafter \@secondoftwo
 \fi
}%
\providecommand \natexlab [1]{#1}%
\providecommand \enquote  [1]{``#1''}%
\providecommand \bibnamefont  [1]{#1}%
\providecommand \bibfnamefont [1]{#1}%
\providecommand \citenamefont [1]{#1}%
\providecommand \href@noop [0]{\@secondoftwo}%
\providecommand \href [0]{\begingroup \@sanitize@url \@href}%
\providecommand \@href[1]{\@@startlink{#1}\@@href}%
\providecommand \@@href[1]{\endgroup#1\@@endlink}%
\providecommand \@sanitize@url [0]{\catcode `\\12\catcode `\$12\catcode
  `\&12\catcode `\#12\catcode `\^12\catcode `\_12\catcode `\%12\relax}%
\providecommand \@@startlink[1]{}%
\providecommand \@@endlink[0]{}%
\providecommand \url  [0]{\begingroup\@sanitize@url \@url }%
\providecommand \@url [1]{\endgroup\@href {#1}{\urlprefix }}%
\providecommand \urlprefix  [0]{URL }%
\providecommand \Eprint [0]{\href }%
\providecommand \doibase [0]{http://dx.doi.org/}%
\providecommand \selectlanguage [0]{\@gobble}%
\providecommand \bibinfo  [0]{\@secondoftwo}%
\providecommand \bibfield  [0]{\@secondoftwo}%
\providecommand \translation [1]{[#1]}%
\providecommand \BibitemOpen [0]{}%
\providecommand \bibitemStop [0]{}%
\providecommand \bibitemNoStop [0]{.\EOS\space}%
\providecommand \EOS [0]{\spacefactor3000\relax}%
\providecommand \BibitemShut  [1]{\csname bibitem#1\endcsname}%
\let\auto@bib@innerbib\@empty
\bibitem [{\citenamefont {Wolff}\ and\ \citenamefont
  {Stephens}(1953)}]{Wolff1953}%
  \BibitemOpen
  \bibfield  {author} {\bibinfo {author} {\bibfnamefont {M.~M.}\ \bibnamefont
  {Wolff}}\ and\ \bibinfo {author} {\bibfnamefont {W.~E.}\ \bibnamefont
  {Stephens}},\ }\href {\doibase 10.1063/1.1770801} {\bibfield  {journal}
  {\bibinfo  {journal} {Rev. Sci. Instrum.}\ }\textbf {\bibinfo {volume}
  {24}},\ \bibinfo {pages} {616} (\bibinfo {year} {1953})}\BibitemShut
  {NoStop}%
\bibitem [{\citenamefont {Wiley}\ and\ \citenamefont
  {McLaren}(1955)}]{Wiley1955}%
  \BibitemOpen
  \bibfield  {author} {\bibinfo {author} {\bibfnamefont {W.~C.}\ \bibnamefont
  {Wiley}}\ and\ \bibinfo {author} {\bibfnamefont {I.~H.}\ \bibnamefont
  {McLaren}},\ }\href@noop {} {\bibfield  {journal} {\bibinfo  {journal} {Rev.
  Sci. Instrum.}\ }\textbf {\bibinfo {volume} {26}},\ \bibinfo {pages} {1150}
  (\bibinfo {year} {1955})}\BibitemShut {NoStop}%
\bibitem [{\citenamefont {Katzenstein}\ and\ \citenamefont
  {Friedland}(1955)}]{Katzenstein1955}%
  \BibitemOpen
  \bibfield  {author} {\bibinfo {author} {\bibfnamefont {H.~S.}\ \bibnamefont
  {Katzenstein}}\ and\ \bibinfo {author} {\bibfnamefont {S.~S.}\ \bibnamefont
  {Friedland}},\ }\href {\doibase 10.1063/1.1771290} {\bibfield  {journal}
  {\bibinfo  {journal} {Rev. Sci. Instrum.}\ }\textbf {\bibinfo {volume}
  {26}},\ \bibinfo {pages} {324} (\bibinfo {year} {1955})}\BibitemShut
  {NoStop}%
\bibitem [{\citenamefont {Mamyrin}\ \emph {et~al.}(1973)\citenamefont
  {Mamyrin}, \citenamefont {Karataev}, \citenamefont {Shmikk},\ and\
  \citenamefont {Zagulin}}]{Mamyrin1973}%
  \BibitemOpen
  \bibfield  {author} {\bibinfo {author} {\bibfnamefont {B.~A.}\ \bibnamefont
  {Mamyrin}}, \bibinfo {author} {\bibfnamefont {V.~I.}\ \bibnamefont
  {Karataev}}, \bibinfo {author} {\bibfnamefont {D.~V.}\ \bibnamefont
  {Shmikk}}, \ and\ \bibinfo {author} {\bibfnamefont {V.~A.}\ \bibnamefont
  {Zagulin}},\ }\href
  {http://inis.iaea.org/search/search.aspx?orig_q=RN:04074573} {\ \textbf
  {\bibinfo {volume} {64}},\ \bibinfo {pages} {82} (\bibinfo {year}
  {1973})}\BibitemShut {NoStop}%
\bibitem [{\citenamefont {Goudsmit}(1948)}]{Goudsmit1948}%
  \BibitemOpen
  \bibfield  {author} {\bibinfo {author} {\bibfnamefont {S.~A.}\ \bibnamefont
  {Goudsmit}},\ }\href {\doibase 10.1103/PhysRev.74.622} {\bibfield  {journal}
  {\bibinfo  {journal} {Phys. Rev.}\ }\textbf {\bibinfo {volume} {74}},\
  \bibinfo {pages} {622} (\bibinfo {year} {1948})}\BibitemShut {NoStop}%
\bibitem [{\citenamefont {Hays}\ \emph {et~al.}(1951)\citenamefont {Hays},
  \citenamefont {Richards},\ and\ \citenamefont {Goudsmit}}]{Hays1951}%
  \BibitemOpen
  \bibfield  {author} {\bibinfo {author} {\bibfnamefont {E.~E.}\ \bibnamefont
  {Hays}}, \bibinfo {author} {\bibfnamefont {P.~I.}\ \bibnamefont {Richards}},
  \ and\ \bibinfo {author} {\bibfnamefont {S.~A.}\ \bibnamefont {Goudsmit}},\
  }\href {\doibase 10.1103/PhysRev.84.824} {\bibfield  {journal} {\bibinfo
  {journal} {Phys. Rev.}\ }\textbf {\bibinfo {volume} {84}},\ \bibinfo {pages}
  {824} (\bibinfo {year} {1951})}\BibitemShut {NoStop}%
\bibitem [{\citenamefont {Mamyrin}(2001)}]{MAMYRIN2001251}%
  \BibitemOpen
  \bibfield  {author} {\bibinfo {author} {\bibfnamefont {B.}~\bibnamefont
  {Mamyrin}},\ }\href {\doibase https://doi.org/10.1016/S1387-3806(00)00392-4}
  {\bibfield  {journal} {\bibinfo  {journal} {Int. J. Mass Spectrom.}\ }\textbf
  {\bibinfo {volume} {206}},\ \bibinfo {pages} {251 } (\bibinfo {year}
  {2001})}\BibitemShut {NoStop}%
\bibitem [{\citenamefont {Enke}\ and\ \citenamefont {Dobson}(2007)}]{Enke2007}%
  \BibitemOpen
  \bibfield  {author} {\bibinfo {author} {\bibfnamefont {C.~G.}\ \bibnamefont
  {Enke}}\ and\ \bibinfo {author} {\bibfnamefont {G.~S.}\ \bibnamefont
  {Dobson}},\ }\href {\doibase 10.1021/ac070638u} {\bibfield  {journal}
  {\bibinfo  {journal} {Anal. Chem.}\ }\textbf {\bibinfo {volume} {79}},\
  \bibinfo {pages} {8650} (\bibinfo {year} {2007})}\BibitemShut {NoStop}%
\bibitem [{\citenamefont {Vestal}\ \emph {et~al.}(1995)\citenamefont {Vestal},
  \citenamefont {Juhasz},\ and\ \citenamefont {Martin}}]{Vestal1995}%
  \BibitemOpen
  \bibfield  {author} {\bibinfo {author} {\bibfnamefont {M.~L.}\ \bibnamefont
  {Vestal}}, \bibinfo {author} {\bibfnamefont {P.}~\bibnamefont {Juhasz}}, \
  and\ \bibinfo {author} {\bibfnamefont {S.~A.}\ \bibnamefont {Martin}},\
  }\href {\doibase 10.1002/rcm.1290091115} {\bibfield  {journal} {\bibinfo
  {journal} {Rapid Commun. Mass Spectrom.}\ }\textbf {\bibinfo {volume} {9}},\
  \bibinfo {pages} {1044} (\bibinfo {year} {1995})}\BibitemShut {NoStop}%
\bibitem [{\citenamefont {Takach}\ \emph {et~al.}(1997)\citenamefont {Takach},
  \citenamefont {Hines}, \citenamefont {Patterson}, \citenamefont {Juhasz},
  \citenamefont {Falick}, \citenamefont {Vestal},\ and\ \citenamefont
  {Martin}}]{Takach1997}%
  \BibitemOpen
  \bibfield  {author} {\bibinfo {author} {\bibfnamefont {E.~J.}\ \bibnamefont
  {Takach}}, \bibinfo {author} {\bibfnamefont {W.~M.}\ \bibnamefont {Hines}},
  \bibinfo {author} {\bibfnamefont {D.~H.}\ \bibnamefont {Patterson}}, \bibinfo
  {author} {\bibfnamefont {P.}~\bibnamefont {Juhasz}}, \bibinfo {author}
  {\bibfnamefont {A.~M.}\ \bibnamefont {Falick}}, \bibinfo {author}
  {\bibfnamefont {M.~L.}\ \bibnamefont {Vestal}}, \ and\ \bibinfo {author}
  {\bibfnamefont {S.~A.}\ \bibnamefont {Martin}},\ }\href {\doibase
  10.1023/A:1026376403468} {\bibfield  {journal} {\bibinfo  {journal} {J. of
  Protein Chem.}\ }\textbf {\bibinfo {volume} {16}},\ \bibinfo {pages} {363}
  (\bibinfo {year} {1997})}\BibitemShut {NoStop}%
\bibitem [{\citenamefont {Michael}\ \emph {et~al.}(1992)\citenamefont
  {Michael}, \citenamefont {Chien},\ and\ \citenamefont {Lubman}}]{Steven1992}%
  \BibitemOpen
  \bibfield  {author} {\bibinfo {author} {\bibfnamefont {S.~M.}\ \bibnamefont
  {Michael}}, \bibinfo {author} {\bibfnamefont {M.}~\bibnamefont {Chien}}, \
  and\ \bibinfo {author} {\bibfnamefont {D.~M.}\ \bibnamefont {Lubman}},\
  }\href {\doibase 10.1063/1.1143725} {\bibfield  {journal} {\bibinfo
  {journal} {Rev. Sci. Instrum.}\ }\textbf {\bibinfo {volume} {63}},\ \bibinfo
  {pages} {4277} (\bibinfo {year} {1992})}\BibitemShut {NoStop}%
\bibitem [{\citenamefont {Wollnik}\ and\ \citenamefont
  {Przewloka}(1990)}]{WOLLNIK1990267}%
  \BibitemOpen
  \bibfield  {author} {\bibinfo {author} {\bibfnamefont {H.}~\bibnamefont
  {Wollnik}}\ and\ \bibinfo {author} {\bibfnamefont {M.}~\bibnamefont
  {Przewloka}},\ }\href {\doibase https://doi.org/10.1016/0168-1176(90)85127-N}
  {\bibfield  {journal} {\bibinfo  {journal} {Int. J. Mass Spectrom. Ion
  Proc.}\ }\textbf {\bibinfo {volume} {96}},\ \bibinfo {pages} {267 } (\bibinfo
  {year} {1990})}\BibitemShut {NoStop}%
\bibitem [{\citenamefont {Wienholtz}\ \emph {et~al.}(2013)\citenamefont
  {Wienholtz}, \citenamefont {Beck}, \citenamefont {Blaum}, \citenamefont
  {Borgmann}, \citenamefont {Breitenfeldt}, \citenamefont {Cakirli},
  \citenamefont {George}, \citenamefont {Herfurth}, \citenamefont {Holt},
  \citenamefont {Kowalska}, \citenamefont {Kreim}, \citenamefont {Lunney},
  \citenamefont {Manea}, \citenamefont {Men\`endez}, \citenamefont {Neidherr},
  \citenamefont {Rosenbusch}, \citenamefont {Schweikhard}, \citenamefont
  {Schwenk}, \citenamefont {Simonis}, \citenamefont {Stanja}, \citenamefont
  {Wolf},\ and\ \citenamefont {Zuber}}]{Wienholtz2013}%
  \BibitemOpen
  \bibfield  {author} {\bibinfo {author} {\bibfnamefont {F.}~\bibnamefont
  {Wienholtz}}, \bibinfo {author} {\bibfnamefont {D.}~\bibnamefont {Beck}},
  \bibinfo {author} {\bibfnamefont {K.}~\bibnamefont {Blaum}}, \bibinfo
  {author} {\bibfnamefont {C.}~\bibnamefont {Borgmann}}, \bibinfo {author}
  {\bibfnamefont {M.}~\bibnamefont {Breitenfeldt}}, \bibinfo {author}
  {\bibfnamefont {R.~B.}\ \bibnamefont {Cakirli}}, \bibinfo {author}
  {\bibfnamefont {S.}~\bibnamefont {George}}, \bibinfo {author} {\bibfnamefont
  {F.}~\bibnamefont {Herfurth}}, \bibinfo {author} {\bibfnamefont {J.~D.}\
  \bibnamefont {Holt}}, \bibinfo {author} {\bibfnamefont {M.}~\bibnamefont
  {Kowalska}}, \bibinfo {author} {\bibfnamefont {S.}~\bibnamefont {Kreim}},
  \bibinfo {author} {\bibfnamefont {D.}~\bibnamefont {Lunney}}, \bibinfo
  {author} {\bibfnamefont {V.}~\bibnamefont {Manea}}, \bibinfo {author}
  {\bibfnamefont {J.}~\bibnamefont {Men\`endez}}, \bibinfo {author}
  {\bibfnamefont {D.}~\bibnamefont {Neidherr}}, \bibinfo {author}
  {\bibfnamefont {M.}~\bibnamefont {Rosenbusch}}, \bibinfo {author}
  {\bibfnamefont {L.}~\bibnamefont {Schweikhard}}, \bibinfo {author}
  {\bibfnamefont {A.}~\bibnamefont {Schwenk}}, \bibinfo {author} {\bibfnamefont
  {J.}~\bibnamefont {Simonis}}, \bibinfo {author} {\bibfnamefont
  {J.}~\bibnamefont {Stanja}}, \bibinfo {author} {\bibfnamefont {R.~N.}\
  \bibnamefont {Wolf}}, \ and\ \bibinfo {author} {\bibfnamefont
  {K.}~\bibnamefont {Zuber}},\ }\href {http://dx.doi.org/10.1038/nature12226}
  {\bibfield  {journal} {\bibinfo  {journal} {Nature}\ }\textbf {\bibinfo
  {volume} {498}},\ \bibinfo {pages} {346} (\bibinfo {year}
  {2013})}\BibitemShut {NoStop}%
\bibitem [{\citenamefont {Ito}\ \emph {et~al.}(2013)\citenamefont {Ito},
  \citenamefont {Schury}, \citenamefont {Wada}, \citenamefont {Naimi},
  \citenamefont {Sonoda}, \citenamefont {Mita}, \citenamefont {Arai},
  \citenamefont {Takamine}, \citenamefont {Okada}, \citenamefont {Ozawa},\ and\
  \citenamefont {Wollnik}}]{ITO2013}%
  \BibitemOpen
  \bibfield  {author} {\bibinfo {author} {\bibfnamefont {Y.}~\bibnamefont
  {Ito}}, \bibinfo {author} {\bibfnamefont {P.}~\bibnamefont {Schury}},
  \bibinfo {author} {\bibfnamefont {M.}~\bibnamefont {Wada}}, \bibinfo {author}
  {\bibfnamefont {S.}~\bibnamefont {Naimi}}, \bibinfo {author} {\bibfnamefont
  {T.}~\bibnamefont {Sonoda}}, \bibinfo {author} {\bibfnamefont
  {H.}~\bibnamefont {Mita}}, \bibinfo {author} {\bibfnamefont {F.}~\bibnamefont
  {Arai}}, \bibinfo {author} {\bibfnamefont {A.}~\bibnamefont {Takamine}},
  \bibinfo {author} {\bibfnamefont {K.}~\bibnamefont {Okada}}, \bibinfo
  {author} {\bibfnamefont {A.}~\bibnamefont {Ozawa}}, \ and\ \bibinfo {author}
  {\bibfnamefont {H.}~\bibnamefont {Wollnik}},\ }\href {\doibase
  10.1103/PhysRevC.88.011306} {\bibfield  {journal} {\bibinfo  {journal} {Phys.
  Rev. C}\ }\textbf {\bibinfo {volume} {88}},\ \bibinfo {pages} {011306}
  (\bibinfo {year} {2013})}\BibitemShut {NoStop}%
\bibitem [{\citenamefont {Wolf}\ \emph {et~al.}(2013)\citenamefont {Wolf},
  \citenamefont {Wienholtz}, \citenamefont {Atanasov}, \citenamefont {Beck},
  \citenamefont {Blaum}, \citenamefont {Borgmann}, \citenamefont {Herfurth},
  \citenamefont {Kowalska}, \citenamefont {Kreim}, \citenamefont {Litvinov},
  \citenamefont {Lunney}, \citenamefont {Manea}, \citenamefont {Neidherr},
  \citenamefont {Rosenbusch}, \citenamefont {Schweikhard}, \citenamefont
  {Stanja},\ and\ \citenamefont {Zuber}}]{Wolf2013a}%
  \BibitemOpen
  \bibfield  {author} {\bibinfo {author} {\bibfnamefont {R.}~\bibnamefont
  {Wolf}}, \bibinfo {author} {\bibfnamefont {F.}~\bibnamefont {Wienholtz}},
  \bibinfo {author} {\bibfnamefont {D.}~\bibnamefont {Atanasov}}, \bibinfo
  {author} {\bibfnamefont {D.}~\bibnamefont {Beck}}, \bibinfo {author}
  {\bibfnamefont {K.}~\bibnamefont {Blaum}}, \bibinfo {author} {\bibfnamefont
  {C.}~\bibnamefont {Borgmann}}, \bibinfo {author} {\bibfnamefont
  {F.}~\bibnamefont {Herfurth}}, \bibinfo {author} {\bibfnamefont
  {M.}~\bibnamefont {Kowalska}}, \bibinfo {author} {\bibfnamefont
  {S.}~\bibnamefont {Kreim}}, \bibinfo {author} {\bibfnamefont {Y.~A.}\
  \bibnamefont {Litvinov}}, \bibinfo {author} {\bibfnamefont {D.}~\bibnamefont
  {Lunney}}, \bibinfo {author} {\bibfnamefont {V.}~\bibnamefont {Manea}},
  \bibinfo {author} {\bibfnamefont {D.}~\bibnamefont {Neidherr}}, \bibinfo
  {author} {\bibfnamefont {M.}~\bibnamefont {Rosenbusch}}, \bibinfo {author}
  {\bibfnamefont {L.}~\bibnamefont {Schweikhard}}, \bibinfo {author}
  {\bibfnamefont {J.}~\bibnamefont {Stanja}}, \ and\ \bibinfo {author}
  {\bibfnamefont {K.}~\bibnamefont {Zuber}},\ }\href {\doibase
  10.1016/j.ijms.2013.03.020} {\bibfield  {journal} {\bibinfo  {journal} {Int.
  J. Mass Spectrom.}\ }\textbf {\bibinfo {volume} {349-350}},\ \bibinfo {pages}
  {123 } (\bibinfo {year} {2013})}\BibitemShut {NoStop}%
\bibitem [{\citenamefont {Schury}\ \emph
  {et~al.}(2014{\natexlab{a}})\citenamefont {Schury}, \citenamefont {Wada},
  \citenamefont {Ito}, \citenamefont {Arai}, \citenamefont {Naimi},
  \citenamefont {Sonoda}, \citenamefont {Wollnik}, \citenamefont {Shchepunov},
  \citenamefont {Smorra},\ and\ \citenamefont {Yuan}}]{SCHURY201439}%
  \BibitemOpen
  \bibfield  {author} {\bibinfo {author} {\bibfnamefont {P.}~\bibnamefont
  {Schury}}, \bibinfo {author} {\bibfnamefont {M.}~\bibnamefont {Wada}},
  \bibinfo {author} {\bibfnamefont {Y.}~\bibnamefont {Ito}}, \bibinfo {author}
  {\bibfnamefont {F.}~\bibnamefont {Arai}}, \bibinfo {author} {\bibfnamefont
  {S.}~\bibnamefont {Naimi}}, \bibinfo {author} {\bibfnamefont
  {T.}~\bibnamefont {Sonoda}}, \bibinfo {author} {\bibfnamefont
  {H.}~\bibnamefont {Wollnik}}, \bibinfo {author} {\bibfnamefont
  {V.}~\bibnamefont {Shchepunov}}, \bibinfo {author} {\bibfnamefont
  {C.}~\bibnamefont {Smorra}}, \ and\ \bibinfo {author} {\bibfnamefont
  {C.}~\bibnamefont {Yuan}},\ }\href {\doibase
  https://doi.org/10.1016/j.nimb.2014.05.016} {\bibfield  {journal} {\bibinfo
  {journal} {Nucl. Instr. Meth. B}\ }\textbf {\bibinfo {volume} {335}},\
  \bibinfo {pages} {39 } (\bibinfo {year} {2014}{\natexlab{a}})}\BibitemShut
  {NoStop}%
\bibitem [{\citenamefont {Chauveau}\ \emph {et~al.}(2016)\citenamefont
  {Chauveau}, \citenamefont {Delahaye}, \citenamefont {France}, \citenamefont
  {Abir}, \citenamefont {Lory}, \citenamefont {Merrer}, \citenamefont
  {Rosenbusch}, \citenamefont {Schweikhard},\ and\ \citenamefont
  {Wolf}}]{CHAUVEAU2016211}%
  \BibitemOpen
  \bibfield  {author} {\bibinfo {author} {\bibfnamefont {P.}~\bibnamefont
  {Chauveau}}, \bibinfo {author} {\bibfnamefont {P.}~\bibnamefont {Delahaye}},
  \bibinfo {author} {\bibfnamefont {G.~D.}\ \bibnamefont {France}}, \bibinfo
  {author} {\bibfnamefont {S.~E.}\ \bibnamefont {Abir}}, \bibinfo {author}
  {\bibfnamefont {J.}~\bibnamefont {Lory}}, \bibinfo {author} {\bibfnamefont
  {Y.}~\bibnamefont {Merrer}}, \bibinfo {author} {\bibfnamefont
  {M.}~\bibnamefont {Rosenbusch}}, \bibinfo {author} {\bibfnamefont
  {L.}~\bibnamefont {Schweikhard}}, \ and\ \bibinfo {author} {\bibfnamefont
  {R.}~\bibnamefont {Wolf}},\ }\href {\doibase
  https://doi.org/10.1016/j.nimb.2016.01.025} {\bibfield  {journal} {\bibinfo
  {journal} {Nucl. Instrum. Meth. B}\ }\textbf {\bibinfo {volume} {376}},\
  \bibinfo {pages} {211 } (\bibinfo {year} {2016})},\ \bibinfo {note}
  {proceedings of the XVIIth International Conference on Electromagnetic
  Isotope Separators and Related Topics (EMIS2015), Grand Rapids, MI, U.S.A.,
  11-15 May 2015}\BibitemShut {NoStop}%
\bibitem [{\citenamefont {Hirsh}\ \emph {et~al.}(2016)\citenamefont {Hirsh},
  \citenamefont {Paul}, \citenamefont {Burkey}, \citenamefont {Aprahamian},
  \citenamefont {Buchinger}, \citenamefont {Caldwell}, \citenamefont {Clark},
  \citenamefont {Levand}, \citenamefont {Ying}, \citenamefont {Marley},
  \citenamefont {Morgan}, \citenamefont {Nystrom}, \citenamefont {Orford},
  \citenamefont {Galván}, \citenamefont {Rohrer}, \citenamefont {Savard},
  \citenamefont {Sharma},\ and\ \citenamefont {Siegl}}]{HIRSH2016229}%
  \BibitemOpen
  \bibfield  {author} {\bibinfo {author} {\bibfnamefont {T.~Y.}\ \bibnamefont
  {Hirsh}}, \bibinfo {author} {\bibfnamefont {N.}~\bibnamefont {Paul}},
  \bibinfo {author} {\bibfnamefont {M.}~\bibnamefont {Burkey}}, \bibinfo
  {author} {\bibfnamefont {A.}~\bibnamefont {Aprahamian}}, \bibinfo {author}
  {\bibfnamefont {F.}~\bibnamefont {Buchinger}}, \bibinfo {author}
  {\bibfnamefont {S.}~\bibnamefont {Caldwell}}, \bibinfo {author}
  {\bibfnamefont {J.~A.}\ \bibnamefont {Clark}}, \bibinfo {author}
  {\bibfnamefont {A.~F.}\ \bibnamefont {Levand}}, \bibinfo {author}
  {\bibfnamefont {L.~L.}\ \bibnamefont {Ying}}, \bibinfo {author}
  {\bibfnamefont {S.~T.}\ \bibnamefont {Marley}}, \bibinfo {author}
  {\bibfnamefont {G.~E.}\ \bibnamefont {Morgan}}, \bibinfo {author}
  {\bibfnamefont {A.}~\bibnamefont {Nystrom}}, \bibinfo {author} {\bibfnamefont
  {R.}~\bibnamefont {Orford}}, \bibinfo {author} {\bibfnamefont {A.~P.}\
  \bibnamefont {Galván}}, \bibinfo {author} {\bibfnamefont {J.}~\bibnamefont
  {Rohrer}}, \bibinfo {author} {\bibfnamefont {G.}~\bibnamefont {Savard}},
  \bibinfo {author} {\bibfnamefont {K.~S.}\ \bibnamefont {Sharma}}, \ and\
  \bibinfo {author} {\bibfnamefont {K.}~\bibnamefont {Siegl}},\ }\href
  {\doibase https://doi.org/10.1016/j.nimb.2015.12.037} {\bibfield  {journal}
  {\bibinfo  {journal} {Nucl. Instrum. Meth. B}\ }\textbf {\bibinfo {volume}
  {376}},\ \bibinfo {pages} {229 } (\bibinfo {year} {2016})},\ \bibinfo {note}
  {proceedings of the XVIIth International Conference on Electromagnetic
  Isotope Separators and Related Topics (EMIS2015), Grand Rapids, MI, U.S.A.,
  11-15 May 2015}\BibitemShut {NoStop}%
\bibitem [{\citenamefont {Ito}\ \emph {et~al.}(2018)\citenamefont {Ito},
  \citenamefont {Schury}, \citenamefont {Wada}, \citenamefont {Arai},
  \citenamefont {Haba}, \citenamefont {Hirayama}, \citenamefont {Ishizawa},
  \citenamefont {Kaji}, \citenamefont {Kimura}, \citenamefont {Koura},
  \citenamefont {MacCormick}, \citenamefont {Miyatake}, \citenamefont {Moon},
  \citenamefont {Morimoto}, \citenamefont {Morita}, \citenamefont {Mukai},
  \citenamefont {Murray}, \citenamefont {Niwase}, \citenamefont {Okada},
  \citenamefont {Ozawa}, \citenamefont {Rosenbusch}, \citenamefont {Takamine},
  \citenamefont {Tanaka}, \citenamefont {Watanabe}, \citenamefont {Wollnik},\
  and\ \citenamefont {Yamaki}}]{Ito2017}%
  \BibitemOpen
  \bibfield  {author} {\bibinfo {author} {\bibfnamefont {Y.}~\bibnamefont
  {Ito}}, \bibinfo {author} {\bibfnamefont {P.}~\bibnamefont {Schury}},
  \bibinfo {author} {\bibfnamefont {M.}~\bibnamefont {Wada}}, \bibinfo {author}
  {\bibfnamefont {F.}~\bibnamefont {Arai}}, \bibinfo {author} {\bibfnamefont
  {H.}~\bibnamefont {Haba}}, \bibinfo {author} {\bibfnamefont {Y.}~\bibnamefont
  {Hirayama}}, \bibinfo {author} {\bibfnamefont {S.}~\bibnamefont {Ishizawa}},
  \bibinfo {author} {\bibfnamefont {D.}~\bibnamefont {Kaji}}, \bibinfo {author}
  {\bibfnamefont {S.}~\bibnamefont {Kimura}}, \bibinfo {author} {\bibfnamefont
  {H.}~\bibnamefont {Koura}}, \bibinfo {author} {\bibfnamefont
  {M.}~\bibnamefont {MacCormick}}, \bibinfo {author} {\bibfnamefont
  {H.}~\bibnamefont {Miyatake}}, \bibinfo {author} {\bibfnamefont {J.~Y.}\
  \bibnamefont {Moon}}, \bibinfo {author} {\bibfnamefont {K.}~\bibnamefont
  {Morimoto}}, \bibinfo {author} {\bibfnamefont {K.}~\bibnamefont {Morita}},
  \bibinfo {author} {\bibfnamefont {M.}~\bibnamefont {Mukai}}, \bibinfo
  {author} {\bibfnamefont {I.}~\bibnamefont {Murray}}, \bibinfo {author}
  {\bibfnamefont {T.}~\bibnamefont {Niwase}}, \bibinfo {author} {\bibfnamefont
  {K.}~\bibnamefont {Okada}}, \bibinfo {author} {\bibfnamefont
  {A.}~\bibnamefont {Ozawa}}, \bibinfo {author} {\bibfnamefont
  {M.}~\bibnamefont {Rosenbusch}}, \bibinfo {author} {\bibfnamefont
  {A.}~\bibnamefont {Takamine}}, \bibinfo {author} {\bibfnamefont
  {T.}~\bibnamefont {Tanaka}}, \bibinfo {author} {\bibfnamefont {Y.~X.}\
  \bibnamefont {Watanabe}}, \bibinfo {author} {\bibfnamefont {H.}~\bibnamefont
  {Wollnik}}, \ and\ \bibinfo {author} {\bibfnamefont {S.}~\bibnamefont
  {Yamaki}},\ }\href {\doibase 10.1103/PhysRevLett.120.152501} {\bibfield
  {journal} {\bibinfo  {journal} {Phys. Rev. Lett.}\ }\textbf {\bibinfo
  {volume} {120}},\ \bibinfo {pages} {152501} (\bibinfo {year}
  {2018})}\BibitemShut {NoStop}%
\bibitem [{\citenamefont {Jesch}\ \emph {et~al.}(2017)\citenamefont {Jesch},
  \citenamefont {Dickel}, \citenamefont {Pla{\ss}}, \citenamefont {Short},
  \citenamefont {Andres}, \citenamefont {Dilling}, \citenamefont {Geissel},
  \citenamefont {Greiner}, \citenamefont {Lang}, \citenamefont {Leach},
  \citenamefont {Lippert}, \citenamefont {Scheidenberger},\ and\ \citenamefont
  {Yavor}}]{Jesh2017}%
  \BibitemOpen
  \bibfield  {author} {\bibinfo {author} {\bibfnamefont {C.}~\bibnamefont
  {Jesch}}, \bibinfo {author} {\bibfnamefont {T.}~\bibnamefont {Dickel}},
  \bibinfo {author} {\bibfnamefont {W.~R.}\ \bibnamefont {Pla{\ss}}}, \bibinfo
  {author} {\bibfnamefont {D.}~\bibnamefont {Short}}, \bibinfo {author}
  {\bibfnamefont {S.~A.~S.}\ \bibnamefont {Andres}}, \bibinfo {author}
  {\bibfnamefont {J.}~\bibnamefont {Dilling}}, \bibinfo {author} {\bibfnamefont
  {H.}~\bibnamefont {Geissel}}, \bibinfo {author} {\bibfnamefont
  {F.}~\bibnamefont {Greiner}}, \bibinfo {author} {\bibfnamefont
  {J.}~\bibnamefont {Lang}}, \bibinfo {author} {\bibfnamefont {K.~G.}\
  \bibnamefont {Leach}}, \bibinfo {author} {\bibfnamefont {W.}~\bibnamefont
  {Lippert}}, \bibinfo {author} {\bibfnamefont {C.}~\bibnamefont
  {Scheidenberger}}, \ and\ \bibinfo {author} {\bibfnamefont {M.~I.}\
  \bibnamefont {Yavor}},\ }in\ \href@noop {} {\emph {\bibinfo {booktitle} {TCP
  2014}}},\ \bibinfo {editor} {edited by\ \bibinfo {editor} {\bibfnamefont
  {M.}~\bibnamefont {Wada}}, \bibinfo {editor} {\bibfnamefont {P.}~\bibnamefont
  {Schury}}, \ and\ \bibinfo {editor} {\bibfnamefont {Y.}~\bibnamefont
  {Ichikawa}}}\ (\bibinfo  {publisher} {Springer International Publishing},\
  \bibinfo {address} {Cham},\ \bibinfo {year} {2017})\ pp.\ \bibinfo {pages}
  {175--184}\BibitemShut {NoStop}%
\bibitem [{\citenamefont {Ayet San~Andr\'es}\ \emph {et~al.}(2019)\citenamefont
  {Ayet San~Andr\'es}, \citenamefont {Hornung}, \citenamefont {Ebert},
  \citenamefont {Pla\ss{}}, \citenamefont {Dickel}, \citenamefont {Geissel},
  \citenamefont {Scheidenberger}, \citenamefont {Bergmann}, \citenamefont
  {Greiner}, \citenamefont {Haettner}, \citenamefont {Jesch}, \citenamefont
  {Lippert}, \citenamefont {Mardor}, \citenamefont {Miskun}, \citenamefont
  {Patyk}, \citenamefont {Pietri}, \citenamefont {Pihktelev}, \citenamefont
  {Purushothaman}, \citenamefont {Reiter}, \citenamefont {Rink}, \citenamefont
  {Weick}, \citenamefont {Yavor}, \citenamefont {Bagchi}, \citenamefont
  {Charviakova}, \citenamefont {Constantin}, \citenamefont {Diwisch},
  \citenamefont {Finlay}, \citenamefont {Kaur}, \citenamefont {Kn\"obel},
  \citenamefont {Lang}, \citenamefont {Mei}, \citenamefont {Moore},
  \citenamefont {Otto}, \citenamefont {Pohjalainen}, \citenamefont {Prochazka},
  \citenamefont {Rappold}, \citenamefont {Takechi}, \citenamefont {Tanaka},
  \citenamefont {Winfield},\ and\ \citenamefont
  {Xu}}]{SanAndresUndChristine2019}%
  \BibitemOpen
  \bibfield  {author} {\bibinfo {author} {\bibfnamefont {S.}~\bibnamefont {Ayet
  San~Andr\'es}}, \bibinfo {author} {\bibfnamefont {C.}~\bibnamefont
  {Hornung}}, \bibinfo {author} {\bibfnamefont {J.}~\bibnamefont {Ebert}},
  \bibinfo {author} {\bibfnamefont {W.~R.}\ \bibnamefont {Pla\ss{}}}, \bibinfo
  {author} {\bibfnamefont {T.}~\bibnamefont {Dickel}}, \bibinfo {author}
  {\bibfnamefont {H.}~\bibnamefont {Geissel}}, \bibinfo {author} {\bibfnamefont
  {C.}~\bibnamefont {Scheidenberger}}, \bibinfo {author} {\bibfnamefont
  {J.}~\bibnamefont {Bergmann}}, \bibinfo {author} {\bibfnamefont
  {F.}~\bibnamefont {Greiner}}, \bibinfo {author} {\bibfnamefont
  {E.}~\bibnamefont {Haettner}}, \bibinfo {author} {\bibfnamefont
  {C.}~\bibnamefont {Jesch}}, \bibinfo {author} {\bibfnamefont
  {W.}~\bibnamefont {Lippert}}, \bibinfo {author} {\bibfnamefont
  {I.}~\bibnamefont {Mardor}}, \bibinfo {author} {\bibfnamefont
  {I.}~\bibnamefont {Miskun}}, \bibinfo {author} {\bibfnamefont
  {Z.}~\bibnamefont {Patyk}}, \bibinfo {author} {\bibfnamefont
  {S.}~\bibnamefont {Pietri}}, \bibinfo {author} {\bibfnamefont
  {A.}~\bibnamefont {Pihktelev}}, \bibinfo {author} {\bibfnamefont
  {S.}~\bibnamefont {Purushothaman}}, \bibinfo {author} {\bibfnamefont {M.~P.}\
  \bibnamefont {Reiter}}, \bibinfo {author} {\bibfnamefont {A.-K.}\
  \bibnamefont {Rink}}, \bibinfo {author} {\bibfnamefont {H.}~\bibnamefont
  {Weick}}, \bibinfo {author} {\bibfnamefont {M.~I.}\ \bibnamefont {Yavor}},
  \bibinfo {author} {\bibfnamefont {S.}~\bibnamefont {Bagchi}}, \bibinfo
  {author} {\bibfnamefont {V.}~\bibnamefont {Charviakova}}, \bibinfo {author}
  {\bibfnamefont {P.}~\bibnamefont {Constantin}}, \bibinfo {author}
  {\bibfnamefont {M.}~\bibnamefont {Diwisch}}, \bibinfo {author} {\bibfnamefont
  {A.}~\bibnamefont {Finlay}}, \bibinfo {author} {\bibfnamefont
  {S.}~\bibnamefont {Kaur}}, \bibinfo {author} {\bibfnamefont {R.}~\bibnamefont
  {Kn\"obel}}, \bibinfo {author} {\bibfnamefont {J.}~\bibnamefont {Lang}},
  \bibinfo {author} {\bibfnamefont {B.}~\bibnamefont {Mei}}, \bibinfo {author}
  {\bibfnamefont {I.~D.}\ \bibnamefont {Moore}}, \bibinfo {author}
  {\bibfnamefont {J.-H.}\ \bibnamefont {Otto}}, \bibinfo {author}
  {\bibfnamefont {I.}~\bibnamefont {Pohjalainen}}, \bibinfo {author}
  {\bibfnamefont {A.}~\bibnamefont {Prochazka}}, \bibinfo {author}
  {\bibfnamefont {C.}~\bibnamefont {Rappold}}, \bibinfo {author} {\bibfnamefont
  {M.}~\bibnamefont {Takechi}}, \bibinfo {author} {\bibfnamefont {Y.~K.}\
  \bibnamefont {Tanaka}}, \bibinfo {author} {\bibfnamefont {J.~S.}\
  \bibnamefont {Winfield}}, \ and\ \bibinfo {author} {\bibfnamefont
  {X.}~\bibnamefont {Xu}},\ }\href {\doibase 10.1103/PhysRevC.99.064313}
  {\bibfield  {journal} {\bibinfo  {journal} {Phys. Rev. C}\ }\textbf {\bibinfo
  {volume} {99}},\ \bibinfo {pages} {064313} (\bibinfo {year}
  {2019})}\BibitemShut {NoStop}%
\bibitem [{\citenamefont {Kimura}\ \emph {et~al.}(2018)\citenamefont {Kimura},
  \citenamefont {Ito}, \citenamefont {Kaji}, \citenamefont {Schury},
  \citenamefont {Wada}, \citenamefont {Haba}, \citenamefont {Hashimoto},
  \citenamefont {Hirayama}, \citenamefont {MacCormick}, \citenamefont
  {Miyatake}, \citenamefont {Moon}, \citenamefont {Morimoto}, \citenamefont
  {Mukai}, \citenamefont {Murray}, \citenamefont {Ozawa}, \citenamefont
  {Rosenbusch}, \citenamefont {Schatz}, \citenamefont {Takamine}, \citenamefont
  {Tanaka}, \citenamefont {Watanabe},\ and\ \citenamefont
  {Wollnik}}]{KIMURA2018134}%
  \BibitemOpen
  \bibfield  {author} {\bibinfo {author} {\bibfnamefont {S.}~\bibnamefont
  {Kimura}}, \bibinfo {author} {\bibfnamefont {Y.}~\bibnamefont {Ito}},
  \bibinfo {author} {\bibfnamefont {D.}~\bibnamefont {Kaji}}, \bibinfo {author}
  {\bibfnamefont {P.}~\bibnamefont {Schury}}, \bibinfo {author} {\bibfnamefont
  {M.}~\bibnamefont {Wada}}, \bibinfo {author} {\bibfnamefont {H.}~\bibnamefont
  {Haba}}, \bibinfo {author} {\bibfnamefont {T.}~\bibnamefont {Hashimoto}},
  \bibinfo {author} {\bibfnamefont {Y.}~\bibnamefont {Hirayama}}, \bibinfo
  {author} {\bibfnamefont {M.}~\bibnamefont {MacCormick}}, \bibinfo {author}
  {\bibfnamefont {H.}~\bibnamefont {Miyatake}}, \bibinfo {author}
  {\bibfnamefont {J.}~\bibnamefont {Moon}}, \bibinfo {author} {\bibfnamefont
  {K.}~\bibnamefont {Morimoto}}, \bibinfo {author} {\bibfnamefont
  {M.}~\bibnamefont {Mukai}}, \bibinfo {author} {\bibfnamefont
  {I.}~\bibnamefont {Murray}}, \bibinfo {author} {\bibfnamefont
  {A.}~\bibnamefont {Ozawa}}, \bibinfo {author} {\bibfnamefont
  {M.}~\bibnamefont {Rosenbusch}}, \bibinfo {author} {\bibfnamefont
  {H.}~\bibnamefont {Schatz}}, \bibinfo {author} {\bibfnamefont
  {A.}~\bibnamefont {Takamine}}, \bibinfo {author} {\bibfnamefont
  {T.}~\bibnamefont {Tanaka}}, \bibinfo {author} {\bibfnamefont
  {Y.}~\bibnamefont {Watanabe}}, \ and\ \bibinfo {author} {\bibfnamefont
  {H.}~\bibnamefont {Wollnik}},\ }\href {\doibase
  https://doi.org/10.1016/j.ijms.2018.05.001} {\bibfield  {journal} {\bibinfo
  {journal} {Int. J. Mass Spectrom.}\ }\textbf {\bibinfo {volume} {430}},\
  \bibinfo {pages} {134 } (\bibinfo {year} {2018})}\BibitemShut {NoStop}%
\bibitem [{\citenamefont {Schury}\ \emph
  {et~al.}(2014{\natexlab{b}})\citenamefont {Schury}, \citenamefont {Ito},
  \citenamefont {Wada},\ and\ \citenamefont {Wollnik}}]{SCHURY201419}%
  \BibitemOpen
  \bibfield  {author} {\bibinfo {author} {\bibfnamefont {P.}~\bibnamefont
  {Schury}}, \bibinfo {author} {\bibfnamefont {Y.}~\bibnamefont {Ito}},
  \bibinfo {author} {\bibfnamefont {M.}~\bibnamefont {Wada}}, \ and\ \bibinfo
  {author} {\bibfnamefont {H.}~\bibnamefont {Wollnik}},\ }\href {\doibase
  https://doi.org/10.1016/j.ijms.2013.11.005} {\bibfield  {journal} {\bibinfo
  {journal} {Int. J. Mass Spectrom.}\ }\textbf {\bibinfo {volume} {359}},\
  \bibinfo {pages} {19 } (\bibinfo {year} {2014}{\natexlab{b}})}\BibitemShut
  {NoStop}%
\bibitem [{\citenamefont {Brun}\ and\ \citenamefont
  {Rademakers}(1997)}]{BRUN199781}%
  \BibitemOpen
  \bibfield  {author} {\bibinfo {author} {\bibfnamefont {R.}~\bibnamefont
  {Brun}}\ and\ \bibinfo {author} {\bibfnamefont {F.}~\bibnamefont
  {Rademakers}},\ }\href {\doibase
  https://doi.org/10.1016/S0168-9002(97)00048-X} {\bibfield  {journal}
  {\bibinfo  {journal} {Nucl. Instr. Meth. B}\ }\textbf {\bibinfo {volume}
  {389}},\ \bibinfo {pages} {81 } (\bibinfo {year} {1997})}\BibitemShut
  {NoStop}%
\bibitem [{\citenamefont {Ravn}(1979)}]{RAVN1979201}%
  \BibitemOpen
  \bibfield  {author} {\bibinfo {author} {\bibfnamefont {H.}~\bibnamefont
  {Ravn}},\ }\href {\doibase https://doi.org/10.1016/0370-1573(79)90045-0}
  {\bibfield  {journal} {\bibinfo  {journal} {Phys. Rep.}\ }\textbf {\bibinfo
  {volume} {54}},\ \bibinfo {pages} {201 } (\bibinfo {year}
  {1979})}\BibitemShut {NoStop}%
\bibitem [{\citenamefont {Villari}(2001)}]{VILLARI2001465}%
  \BibitemOpen
  \bibfield  {author} {\bibinfo {author} {\bibfnamefont {A.}~\bibnamefont
  {Villari}},\ }\href {\doibase https://doi.org/10.1016/S0375-9474(01)01107-1}
  {\bibfield  {journal} {\bibinfo  {journal} {Nucl. Phys. A}\ }\textbf
  {\bibinfo {volume} {693}},\ \bibinfo {pages} {465 } (\bibinfo {year}
  {2001})}\BibitemShut {NoStop}%
\bibitem [{\citenamefont {Harss}\ \emph {et~al.}(2000)\citenamefont {Harss},
  \citenamefont {Pardo}, \citenamefont {Rehm}, \citenamefont {Borasi},
  \citenamefont {Greene}, \citenamefont {Janssens}, \citenamefont {Jiang},
  \citenamefont {Nolen}, \citenamefont {Paul}, \citenamefont {Schiffer},
  \citenamefont {Segel}, \citenamefont {Specht}, \citenamefont {Wang},
  \citenamefont {Wilt},\ and\ \citenamefont {Zabransky}}]{Harss2001}%
  \BibitemOpen
  \bibfield  {author} {\bibinfo {author} {\bibfnamefont {B.}~\bibnamefont
  {Harss}}, \bibinfo {author} {\bibfnamefont {R.~C.}\ \bibnamefont {Pardo}},
  \bibinfo {author} {\bibfnamefont {K.~E.}\ \bibnamefont {Rehm}}, \bibinfo
  {author} {\bibfnamefont {F.}~\bibnamefont {Borasi}}, \bibinfo {author}
  {\bibfnamefont {J.~P.}\ \bibnamefont {Greene}}, \bibinfo {author}
  {\bibfnamefont {R.~V.~F.}\ \bibnamefont {Janssens}}, \bibinfo {author}
  {\bibfnamefont {C.~L.}\ \bibnamefont {Jiang}}, \bibinfo {author}
  {\bibfnamefont {J.}~\bibnamefont {Nolen}}, \bibinfo {author} {\bibfnamefont
  {M.}~\bibnamefont {Paul}}, \bibinfo {author} {\bibfnamefont {J.~P.}\
  \bibnamefont {Schiffer}}, \bibinfo {author} {\bibfnamefont {R.~E.}\
  \bibnamefont {Segel}}, \bibinfo {author} {\bibfnamefont {J.}~\bibnamefont
  {Specht}}, \bibinfo {author} {\bibfnamefont {T.~F.}\ \bibnamefont {Wang}},
  \bibinfo {author} {\bibfnamefont {P.}~\bibnamefont {Wilt}}, \ and\ \bibinfo
  {author} {\bibfnamefont {B.}~\bibnamefont {Zabransky}},\ }\href {\doibase
  10.1063/1.1150211} {\bibfield  {journal} {\bibinfo  {journal} {Rev. Sci.
  Instrum.}\ }\textbf {\bibinfo {volume} {71}},\ \bibinfo {pages} {380}
  (\bibinfo {year} {2000})}\BibitemShut {NoStop}%
\bibitem [{\citenamefont {Yano}(2007)}]{YANO20071009}%
  \BibitemOpen
  \bibfield  {author} {\bibinfo {author} {\bibfnamefont {Y.}~\bibnamefont
  {Yano}},\ }\href {\doibase https://doi.org/10.1016/j.nimb.2007.04.174}
  {\bibfield  {journal} {\bibinfo  {journal} {Nucl. Instrum. Meth. B}\ }\textbf
  {\bibinfo {volume} {261}},\ \bibinfo {pages} {1009 } (\bibinfo {year}
  {2007})}\BibitemShut {NoStop}%
\bibitem [{\citenamefont {Naimi}\ \emph {et~al.}(2013)\citenamefont {Naimi},
  \citenamefont {Nakamura}, \citenamefont {Ito}, \citenamefont {Mita},
  \citenamefont {Okada}, \citenamefont {Ozawa}, \citenamefont {Schury},
  \citenamefont {Sonoda}, \citenamefont {Takamine}, \citenamefont {Wada},\ and\
  \citenamefont {Wollnik}}]{NAIMI201324}%
  \BibitemOpen
  \bibfield  {author} {\bibinfo {author} {\bibfnamefont {S.}~\bibnamefont
  {Naimi}}, \bibinfo {author} {\bibfnamefont {S.}~\bibnamefont {Nakamura}},
  \bibinfo {author} {\bibfnamefont {Y.}~\bibnamefont {Ito}}, \bibinfo {author}
  {\bibfnamefont {H.}~\bibnamefont {Mita}}, \bibinfo {author} {\bibfnamefont
  {K.}~\bibnamefont {Okada}}, \bibinfo {author} {\bibfnamefont
  {A.}~\bibnamefont {Ozawa}}, \bibinfo {author} {\bibfnamefont
  {P.}~\bibnamefont {Schury}}, \bibinfo {author} {\bibfnamefont
  {T.}~\bibnamefont {Sonoda}}, \bibinfo {author} {\bibfnamefont
  {A.}~\bibnamefont {Takamine}}, \bibinfo {author} {\bibfnamefont
  {M.}~\bibnamefont {Wada}}, \ and\ \bibinfo {author} {\bibfnamefont
  {H.}~\bibnamefont {Wollnik}},\ }\href {\doibase
  https://doi.org/10.1016/j.ijms.2012.12.009} {\bibfield  {journal} {\bibinfo
  {journal} {International Journal of Mass Spectrometry}\ }\textbf {\bibinfo
  {volume} {337}},\ \bibinfo {pages} {24 } (\bibinfo {year}
  {2013})}\BibitemShut {NoStop}%
\end{thebibliography}
%

\begin{widetext}

\section{Appendix}
\label{sec:appendix}
	\subsection{Derivatives for uncertainty calculation for the case of a known offset time $t_0$ \\(according to Sec.~\ref{sec:numerical_studies})}
\label{sub:uncertainty-calculation_simple_case}
	We consider the case where the offset time and the transition time is measured with methods other than using further reference ions. In the case of using the same data to obtain $\alpha$ and $\beta$ and performing the mass calculation, Eq.~\ref{eq:finalmass} was used (Fig.~\ref{fig:mass-result} b). For the calculation of the uncertainty considering the input masses and the inner derivatives for $\alpha$ and $\beta$, please see the next subsection.
\begin{equation} 
	\frac{\partial m}{\partial \widetilde{t}_\mathrm{D}} = 2 \sqrt{q m} \left( \frac{1}{2\alpha} + \frac{ 1 }{4\alpha^2 } \frac{\widetilde{t}_\mathrm{D}}{\sqrt{\left( \frac{\widetilde{t}_\mathrm{D}}{2\alpha} \right)^2 - \frac{\beta}{\alpha} }} \right)
\end{equation}
\begin{equation}
	\frac{\partial m}{\partial \widetilde{t}_\mathrm{D1}} = 2 \sqrt{q m} \left( -\frac{\widetilde{t}_\mathrm{D} }{2 \alpha^2} \frac{\partial \alpha}{\partial \widetilde{t}_\mathrm{D1}}         -      \left( \frac{ \widetilde{t}_\mathrm{D}^{2}}{4\alpha^3} \frac{\partial \alpha}{\partial \widetilde{t}_\mathrm{D1}}  + \frac{ \frac{\partial \beta}{\partial \widetilde{t}_\mathrm{D1}} \alpha - \beta \frac{\partial \alpha}{\partial \widetilde{t}_\mathrm{D1}}  }{2 \alpha^2} \right ) \frac{1}{\sqrt{\left( \frac{\widetilde{t}_\mathrm{D}}{2\alpha} \right)^2 - \frac{\beta}{\alpha} }} \right )
\end{equation}
\begin{equation}
	\frac{\partial m}{\partial \widetilde{t}_\mathrm{D2}} = 2 \sqrt{q m} \left( -\frac{\widetilde{t}_\mathrm{D} }{2 \alpha^2} \frac{\partial \alpha}{\partial \widetilde{t}_\mathrm{D2}}         -      \left( \frac{ \widetilde{t}_\mathrm{D}^{2}}{4\alpha^3} \frac{\partial \alpha}{\partial \widetilde{t}_\mathrm{D2}}  + \frac{ \frac{\partial \beta}{\partial \widetilde{t}_\mathrm{D2}} \alpha - \beta \frac{\partial \alpha}{\partial \widetilde{t}_\mathrm{D2}}  }{2 \alpha^2} \right ) \frac{1}{\sqrt{\left( \frac{\widetilde{t}_\mathrm{D}}{2\alpha} \right)^2 - \frac{\beta}{\alpha} }} \right )
\end{equation}
	
	\subsection{Derivatives for uncertainty calculation for the case of an externally measured offset time $t_0$ \\(according to Sec.~\ref{sec:precision_measurements})}
\label{sub:uncertainty-calculation}
	We consider the case where the offset time and the transition time is measured with methods other than using further reference ions. Furthermore, the possibility of rescaling the time-of-flight as performed in Sec.~\ref{sec:precision_measurements} will be included. This means that Eq.~\ref{eq:finalmass-extended} is used, where $\alpha$ and $\beta$ are measured using $\widetilde{t}_\mathrm{D1}$ and $\widetilde{t}_\mathrm{D2}$, whereas $\widetilde{t}_\mathrm{D}^{\prime}$ and $\widetilde{t}_\mathrm{D1}^{\prime}$ are obtained from an independent measurement at a later time and the ratio of $\widetilde{t}_\mathrm{D1}^{\prime} / \widetilde{t}_\mathrm{D1}$ is then used for the correction of the analyte ion's flight time. For the calculation of the mass uncertainty, we will list first the outer derivatives of Eq.~\ref{eq:finalmass-extended}, and then the inner derivatives for $\alpha$ and $\beta$.
\begin{equation} 
	\frac{\partial m}{\partial \widetilde{t}_\mathrm{D}^{\prime}} = 2 \sqrt{q m} \left( \frac{\widetilde{t}_\mathrm{D1}}{2\alpha\widetilde{t}_\mathrm{D1}^{\prime}} + \frac{ \widetilde{t}_\mathrm{D1}^{2} }{4\alpha^2 \widetilde{t}_\mathrm{D1}^{\prime 2}} \frac{\widetilde{t}_\mathrm{D}^{\prime}}{\sqrt{\left( \frac{\widetilde{t}_\mathrm{D}^{\prime} \widetilde{t}_\mathrm{D 1}}{2\alpha \widetilde{t}_\mathrm{D 1}^{\prime}} \right)^2 - \frac{\beta}{\alpha} }} \right)
\end{equation}
\begin{equation}
	\frac{\partial m}{\partial \widetilde{t}_\mathrm{D1}^{\prime}} = 2 \sqrt{q m} \left( -\frac{\widetilde{t}_\mathrm{D}^{\prime} \widetilde{t}_\mathrm{D 1}}{2 \alpha \widetilde{t}_\mathrm{D 1}^{\prime 2}}        -       \frac{ \widetilde{t}_\mathrm{D}^{\prime 2} \widetilde{t}_\mathrm{D 1}^{2} }{4\alpha^2 \widetilde{t}_\mathrm{D1}^{\prime 3}} \frac{1}{\sqrt{\left( \frac{\widetilde{t}_\mathrm{D}^{\prime} \widetilde{t}_\mathrm{D 1}}{2\alpha \widetilde{t}_\mathrm{D 1}^{\prime}} \right)^2 - \frac{\beta}{\alpha} }}   \right)
\end{equation}
\begin{equation}
	\frac{\partial m}{\partial \widetilde{t}_\mathrm{D1}} = 2 \sqrt{q m} \left[  \frac{ \widetilde{t}_\mathrm{D }^{\prime} }{2 \widetilde{t}_\mathrm{D 1}^{\prime}} \left( \frac{\alpha - \widetilde{t}_\mathrm{D 1}\frac{\partial \alpha}{\partial \widetilde{t}_\mathrm{D 1}}}{\alpha^2} \right)  + \left( \frac{ \widetilde{t}_\mathrm{D }^{\prime 2} \widetilde{t}_\mathrm{D 1}}{ 4 \alpha \widetilde{t}_\mathrm{D 1}^{\prime 2}}    \left( \frac{\alpha - \widetilde{t}_\mathrm{D 1}\frac{\partial \alpha}{\partial \widetilde{t}_\mathrm{D 1}}}{\alpha^2} \right) - \frac{  \frac{\partial \beta}{\partial \widetilde{t}_\mathrm{D 1}} \alpha - \beta \frac{\partial \alpha}{\partial \widetilde{t}_\mathrm{D 1}}   }{2 \alpha^2}              \right) \frac{1}{\sqrt{\left( \frac{\widetilde{t}_\mathrm{D}^{\prime} \widetilde{t}_\mathrm{D 1}}{2\alpha \widetilde{t}_\mathrm{D 1}^{\prime}} \right)^2 - \frac{\beta}{\alpha} }} \right]
\end{equation}
\begin{equation}
	\frac{\partial m}{\partial \widetilde{t}_\mathrm{D2}} = 2 \sqrt{q m} \left[ - \frac{ \widetilde{t}_\mathrm{D }^{\prime} \widetilde{t}_\mathrm{D 1}}{2 \alpha^2 \widetilde{t}_\mathrm{D 1}^{\prime}} \frac{\partial \alpha}{\partial \widetilde{t}_\mathrm{D 2}}  - \left( \frac{ \widetilde{t}_\mathrm{D }^{\prime 2} \widetilde{t}_\mathrm{D 1}^{2}}{ 4 \alpha^3 \widetilde{t}_\mathrm{D 1}^{\prime 2}}  \frac{\partial \alpha}{\partial \widetilde{t}_\mathrm{D 2}} + \frac{  \frac{\partial \beta}{\partial \widetilde{t}_\mathrm{D 2}} \alpha - \beta \frac{\partial \alpha}{\partial \widetilde{t}_\mathrm{D 2}}   }{2 \alpha^2}              \right) \frac{1}{\sqrt{\left( \frac{\widetilde{t}_\mathrm{D}^{\prime} \widetilde{t}_\mathrm{D 1}}{2\alpha \widetilde{t}_\mathrm{D 1}^{\prime}} \right)^2 - \frac{\beta}{\alpha} }} \right]
\end{equation}
	With $X_1 = \sqrt{m_1/q_1}$ and $X_2 = \sqrt{m_2/q_2}$ for the two input masses used only implicitly in $\alpha$ and $\beta$, the outer derivatives for $X_1$ and $X_2$ equally yield:
\begin{equation}
	\frac{\partial m}{\partial X_{1,2}} = 2 \sqrt{q m}  \left[ - \frac{ \widetilde{t}_\mathrm{D }^{\prime} \widetilde{t}_\mathrm{D 1}}{2 \alpha^2 \widetilde{t}_\mathrm{D1}^{\prime}} \frac{\partial \alpha}{\partial X_{1,2}}  - \left( \frac{ \widetilde{t}_\mathrm{D }^{\prime 2} \widetilde{t}_\mathrm{D 1}^{2}}{ 4 \alpha^3 \widetilde{t}_\mathrm{D 1}^{\prime 2}}  \frac{\partial \alpha}{\partial X_{1,2}} + \frac{  \frac{\partial \beta}{\partial X_{1,2}} \alpha - \beta \frac{\partial \alpha}{\partial X_{1,2}}   }{2 \alpha^2}              \right) \frac{1}{\sqrt{\left( \frac{\widetilde{t}_\mathrm{D}^{\prime} \widetilde{t}_\mathrm{D 1}}{2\alpha \widetilde{t}_\mathrm{D 1}^{\prime}} \right)^2 - \frac{\beta}{\alpha} }} \right] \quad {.}
\end{equation}
The inner derivatives for $\alpha$ and $\beta$ are:
\begin{equation}
	\frac{\partial\alpha}{\partial \widetilde{t}_\mathrm{D1}} = \frac{X1}{X_1^2 - X_2^2}  
\end{equation}
\begin{equation}
	\frac{\partial\alpha}{\partial \widetilde{t}_\mathrm{D2}} = -\frac{X2}{X_1^2 - X_2^2}  
\end{equation}
\begin{equation}
	\frac{\partial\alpha}{\partial X_1} = - \widetilde{t}_\mathrm{D 1} \frac{X_1^2 + X_2^2}{\left( X_1^2 - X_2^2 \right)^2} + \frac{2 \widetilde{t}_\mathrm{D 2} X_1 X_2}{\left( X_1^2 - X_2^2 \right)^2}  
\end{equation}
\begin{equation}
	\frac{\partial\alpha}{\partial X_2} = - \widetilde{t}_\mathrm{D 2} \frac{X_1^2 + X_2^2}{\left( X_1^2 - X_2^2 \right)^2} + \frac{2 \widetilde{t}_\mathrm{D 1} X_1 X_2}{\left( X_1^2 - X_2^2 \right)^2}  
\end{equation}
\begin{equation}
	\frac{\partial\beta}{\partial \widetilde{t}_\mathrm{D1}} = -\frac{X_2^2 X_1}{X_1^2 - X_2^2}
\end{equation}
\begin{equation}
	\frac{\partial\beta}{\partial \widetilde{t}_\mathrm{D2}} = \frac{X_1^2 X_2}{X_1^2 - X_2^2}
\end{equation}
\begin{equation}
	\frac{\partial\beta}{\partial X_1} = \frac{ \widetilde{t}_\mathrm{D 1} X_1^2 X_2^2 - 2 \widetilde{t}_\mathrm{D 2} X_1 X_2^3 + \widetilde{t}_\mathrm{D 1} X_2^4}{\left( X_1^2 - X_2^2 \right)^2}
\end{equation}
\begin{equation}
	\frac{\partial\beta}{\partial X_2} = \frac{ \widetilde{t}_\mathrm{D 2} X_1^4 - 2 \widetilde{t}_\mathrm{D 1} X_1^3 X_2 + \widetilde{t}_\mathrm{D 2} X_1^2 X_2^2}{\left( X_1^2 - X_2^2 \right)^2} \quad {.}
\end{equation}
As the uncertainty of the charge is negligible, it is sufficient to use only the measurement uncertainty of the input masses if necessary.
\begin{equation}
	\frac{\partial X_{1,2}}{\partial m_{1,2}} = \frac{1}{2\sqrt{q_{1,2} m_{1,2}}} 
\end{equation}
Finally, with $ \widetilde{t}_\mathrm{D,D1,D2} = t_\mathrm{D,D1,D2} - t_\mathrm{T}/2 - t_0$, the contribution of the offset time $t_0$ and the transition time $ t_\mathrm{T}$ is:
\begin{equation}
	\frac{\partial \widetilde{m}}{\partial t_\mathrm{T}} = - \frac{1}{2} \left( \frac{\partial m }{\partial \widetilde{t}_\mathrm{D }^{\prime}} + \frac{\partial m }{\partial \widetilde{t}_\mathrm{D 1}^{\prime}} + \frac{\partial m }{\partial \widetilde{t}_\mathrm{D 1}} + \frac{\partial m }{\partial \widetilde{t}_\mathrm{D 2}}\right)
\end{equation}
\begin{equation}
	\frac{\partial \widetilde{m}}{\partial t_0} = - \left( \frac{\partial m }{\partial \widetilde{t}_\mathrm{D }^{\prime}} + \frac{\partial m }{\partial \widetilde{t}_\mathrm{D 1}^{\prime}} + \frac{\partial m }{\partial \widetilde{t}_\mathrm{D 1}} + \frac{\partial m }{\partial \widetilde{t}_\mathrm{D 2}}\right) \quad {.}
\end{equation}

\end{widetext}
\end{document}